\documentclass{article}

\usepackage{PRIMEarxiv}
\usepackage{adjustbox}
\usepackage{caption, subcaption}
\usepackage[section]{placeins}
\usepackage{natbib}
\usepackage[utf8]{inputenc} 
\usepackage[T1]{fontenc}    
\usepackage{hyperref}       
\usepackage{url}            
\usepackage{booktabs}       
\usepackage{amsfonts}       
\usepackage{nicefrac}       
\usepackage{microtype}      
\usepackage{lipsum}
\usepackage{fancyhdr}       
\usepackage{graphicx}       

\hypersetup{
    colorlinks=false,
    pdfborder={0 0 0},
    pdfborderstyle={/S/U/W 0},
}

\title{The Impact of Cloaking Digital Footprints on User Privacy and Personalization}

\author{ \href{https://orcid.org/0000-0003-3784-826X}{\includegraphics[scale=0.06]{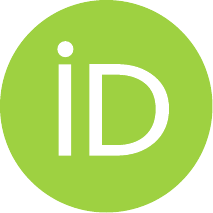}\hspace{1mm}Sofie~Goethals} \\
	University of Antwerp\\
	Antwerp, Belgium \\
	\texttt{sofie.goethals@uantwerpen.be} \\
 \And
        Sandra Matz\\
        Columbia Business \& Engineering Schools\\
        New York, USA\\
    \And
        Foster Provost\\
        NYU Stern\\
       New YOrk, USA\\
    \And
        Yanou Ramon\\
	University of Antwerp\\
	Antwerp, Belgium \\
	\And
        David Martens\\
	University of Antwerp\\
	Antwerp, Belgium \\
	}

\begin{document}
\maketitle

\begin{abstract}
Our online lives generate a wealth of behavioral records -\emph{``digital footprints"}- which are stored and leveraged by technology platforms. This data can be used to create value for users by personalizing services. At the same time, however, it also poses a threat to people's privacy by offering a highly intimate window into their private traits (e.g., their personality, political ideology, sexual orientation). Prior work has proposed a potential remedy: The cloaking of users' footprints. That is, platforms could allow users to hide portions of their digital footprints from predictive algorithms to avoid undesired inferences. While such an approach has been shown to offer privacy protection in the moment, there are two open questions. First, it remains unclear how well cloaking performs over time. As people constantly leave new digital footprints, the algorithm might regain the ability to predict previously cloaked traits.  Second, cloaking digital footprints to avoid one undesirable inference may degrade the performance of models for other, desirable inferences (e.g., those driving desired personalized content). In the light of these research gaps, our contributions are twofold: 1) We propose a novel cloaking strategy that conceals "metafeatures" (automatically generated higher-level categories) and compares its effectiveness against existing cloaking approaches, and 2) we test the spill-over effects of cloaking one trait on the accuracy of inferences on other traits. 
A key finding is that the effectiveness of cloaking degrades over times, but the rate at which it degrades is significantly smaller when cloaking metafeatures rather than individual footprints.
In addition, our findings reveal the expected trade-off between privacy and personalization: Cloaking an undesired trait also partially conceals other desirable traits. 
\end{abstract}



\maketitle

\section{Introduction and related work}
A growing portion of our life happens online: We shop on Amazon, entertain ourselves on Netflix or Spotify and communicate with friend and family via Facebook or Whatsapp. Whether we like it or not, the digital traces we generate during these interactions provide the mediating platforms with an extensive and comprehensive picture of our personal habits and preferences~\citep{matz2020privacy, kosinski2013private}. In fact, research has shown that a person's digital footprints - including their Facebook Likes and status updates, smartphone records or credit card spending - can be used to predict highly intimate characteristics such as sexual or political orientation, personality traits, mental health, or religious views ~\citep{kosinski2013private, matz2017psychological}. Given that most individuals consider these characteristics deeply private, the automated predictions of such traits without individuals knowledge or consent raises important concerns related to people's right to privacy and self-determination ~\citep{matz2020privacy}. The act of drawing highly intimate inferences from seemingly innocuous data, for example, can be regarded as a intrusion of privacy, especially when individuals are neither aware of such inferences being made nor able to object to them. Moreover, the psychological insights platforms (and other third parties) can glean from digital footprints allow them to influence their users' behaviors and decisions through mechanisms of personalization (an approach known as psychological targeting ~\citep{matz2017psychological}).

On the one hand, such personalization approaches might be appreciated by consumers who receive more relevant products and services as a result of targeted advertising and product design~\citep{tran2017personalized}. On the other hand, the ability to predict people's intimate traits and influence their behavior raises serious concerns for individuals and society at large. In countries where homosexuality is illegal, for example, the ability to infer sexual orientation from Facebook Likes could become a death sentence ~\citep{cabanas2018facebook}. Similarly, health assurance companies could attempt to identify people with unhealthy habits or specific health problems, resulting in higher premiums or even rejection of coverage altogether~\citep{cabanas2018facebook}. The perhaps most well-known case of such an abuse is that of Cambridge Analytica, the UK-based PR firm which claimed to have interfered in the 2016 US presidential election by targeting voters with psychologically-tailored advertising on Facebook ~\citep{appel2021psychological}.

Given the seriousness of these potential transgressions and privacy violations by more malicious actors, scientists, activists and policy makers have pushed for legislation that 
aims to prohibit the prediction of protected categories, such as race or religion. Facebook, for example, has faced years of criticism for offering advertisers `interest' categories that have led to the exclusion of people of color from housing ads, fueled political polarization, and helped Big Pharma track users with specific illnesses~\citep{markup2022sensitive,propublica2016housing,wired2019political,citizen2021pharma}. In 2022, Facebook responded to the growing public pressure and changing regulatory landscape by removing the option to target users explicitly based on potentially sensitive traits such as health, race, sexual or political orientation, and religious beliefs~\citep{markup2022sensitive}.\footnote{Some of the interest categories that will be no longer available include `Gay Pride', `Islamic Calendar', and `Lung Cancer Awareness'~\citep{silberlingtech2021, markup2022sensitive}. A comprehensive list of removed Facebook pages can be found here: \url{https://www.propublica.org/article/facebook-lets-advertisers-exclude-users-by-race}.}  

However, as the non-profit news organization The Markup reported, Facebook's attempts at better protecting their users' privacy and preventing discrimination were only partially successful. For example, although \emph{Hispanic Culture} was removed from the target categories available to advertisers, \emph{Spanish Language} was not~\citep{markup2022sensitive}. Although Facebook has since removed additional interests and categories related to protected traits, we argue that playing whack-a-mole across many millions of pages and categories is destined to fail. This is partially the case because few users for which a protected trait is predicted, will actually like pages that explicitly reveal these traits. For example, less than 5\% of users predicted as homosexual were connected with explicitly homosexual pages such as \emph{No H8 Campaign}, \emph{Being Gay} or \emph{I Love Being Gay}~\citep{kosinski2013private}.  

Consequently, the mere act of eliminating certain prediction categories from the platforms prediction or targeting engines is insufficient. This is particularly true when ads or content are still targeted based on data-driven prediction models (e.g., inferring psychological traits) rather than individual interests.  Even if \emph{Gay Pride} is removed as an explicit targeting category and \emph{No H8 Campaign} is removed as a data item for prediction, algorithms may still learn to use other digital footprints to target content or ads that would appeal to gay individuals.


In order to seriously limit the predictability and use of sensitive traits across all users, platforms would have to ban an unreasonable number of pages from their inference algorithms, among them many seemingly neutral pages that will be hard to justify and would likely evoke concerns related to freedom of speech and expression. Moreover, implementing such paternalistic measures may undermine an individual's agency over what they choose to reveal about themselves. For example, should we force users to restrict their online identities if they feel perfectly safe and comfortable about their lifestyle and sexual orientation, and would be delighted to receive advertisements about Drag Shows? In addition, generic one-size-fits-all approaches to restricting certain aspects of online behavior can have negative consequences for socially relevant causes that would benefit from personalization and civic engagement: climate activists and medical researchers, for example, have pointed out that the changes to Facebook's targeting platform have severely limited their ability to reach relevant audiences~\citep{markup2022sensitive}.

In this paper, we examine a more individualized approach that offers users more control and transparency over their online identities, and can be tailored to and by the individual: cloaking certain digital traces that are relevant for the prediction of a particular individual. \citet{chen2017enhancing} propose a “cloaking device" that reveals to users the digital footprints without which the prediction model would not have made the inference and allows them to restrict inference procedures from using them. Let us consider a digital footprint to be a specific aspect of online behavior that is stored about the individual on a technology platform, such as a particular song listened to on Spotify, or a specific page “liked" on Facebook.  \emph{Cloaking} a digital footprint means removing it from the set of data considered by an algorithm drawing inferences.  In the common case of a machine learning model in an AI inference system, where the digital footprints are the features used by the model, cloaking the digital footprint represented by feature $x$ for user $u$ would mean setting the value of $x$ to whatever would be the value if the system had not saved that digital footprint for that user (for example, setting the feature value to zero as an indication that the user did not like the page in question).  This cloaking could be implemented by the platform, by providing users with the option to choose which inferences to avoid.  Alternatively, such transparency could guide users to better decide which data they feel comfortable sharing in the first place; however, for many systems, the digital footprints are the result of simply using the system, so such self-cloaking would involve restricting one's own behavior (e.g., consider songs listened to on Spotify as digital footprints).

The reason why we (and previous authors) focus on cloaking the underlying features, and not the particular inferences, is because the latter cannot protect the user from closely-related inferences in the future~\citep{chen2017enhancing}. This is in line with the current advertising options on Facebook: as mentioned before, it is no longer possible to target people based on certain private traits (e.g., sexual orientation). Hence, advertisers have to rely on associated interests (e.g. Facebook likes that have been empirically shown to be related to a certain sexual orientation) if they want to target a specific private group. The previously proposed cloaking strategy has been shown to be effective in avoiding inferences at the time the cloaking takes place, with relatively little burden on the users.  However, prior work has not investigated how effective the cloaking would be over time, as users continue to leave new digital traces over time.

Digital footprints are usually high-dimensional, sparse, fine-grained behavioral data, for which models normally draw on a large combination of different features as evidence for a possible inference\cite{ramon2021can}.  Therefore, as our results indicate, using a cloaking strategy solely based on singular fine-grained features will not be sufficient, as people will continue to live their lives and behave similarly in the future~\citep{de2020benchmarking,chen2017enhancing}.
For example, our analysis reveals that when we take a snapshot at a later moment, more than 80\% of the people whom the models would target as republican will be subject to targeting again in the future despite cloaking based on the fine-grained features. 
We investigate methods to enhance the longer-term efficacy of cloaking, by grouping the fine-grained features into metafeatures (higher-level feature representations) and cloaking these metafeatures instead.  Our results show that this approach increases the longer-term effectiveness of cloaking considerably and thus enhances desired the privacy protection over time.

Importantly, the implications of cloaking digital footprints to reduce undesired inferences are not uniformly positive. As we discussed at the beginning of this paper, the same digital footprints may be used for different inferences and ultimate purposes. For example, a particular footprint might reveal not only sexual orientation but also the personality trait of Openness. While a user might be concerned about their data being used to predict their sexual orientation and subsequently discriminate against them, they might be appreciative of personalized services and ads that account for their level of openness to experience.  That is, the same traces and mechanisms that may lead to discrimination, can also benefit users in the form of personalized content (e.g.,   individualized playlists, more relevant news etc.). This desired form of personalization not only leads to happier users but also  higher engagement~\citep{fernandez2023playlist} and ultimately higher platform revenue\citep{johnson2020consumer}. 

In this paper we study the potential unintended consequences of cloaking. That is, when users decide to suppress certain digital footprints via cloaking mechanisms, the data available for other desired personalization tasks decreases, potentially reducing their accuracy and effectiveness via spill-over effects.  
To explore this privacy-personalization trade-off, we evaluate the impact of the two previously outlined cloaking strategies (using fine-grained features versus metafeatures) on the accuracy of unrelated prediction tasks that are not subject to cloaking. For example, we examine how cloaking for sexual orientation impacts the predictive performance of a model predicting personality using the same set of digital footprints. Insights into the nature of this trade-off are crucial to empower users to make informed decisions about their online activity and about where on the privacy-personalization trade-off they want to be.

In summary, our study offers three major contributions to the existing literature:
\begin{itemize}
    \item We assess the longer-term effectiveness of cloaking digital footprints, measuring the percentage of targeted individuals whose privacy remains protected over time.  The results show that the effectiveness of cloaking fine-grained features decreases steadily and markedly over time for most inference tasks.
    \item We introduce a new type of cloaking strategy based on metafeatures, and show that it enhances longer-term cloaking protection (as intended).
    \item We examine the privacy-personalization trade-off inherent in using cloaking to protect against unwanted inferences.  Specifically, we show that cloaking for one task can affect the predictive performance of other personalization tasks.  Moreover, the metafeature-based strategies affect other tasks more, highlighting the trade-offs faced by users: better longer-term privacy protection indeed can reduce desired personalization performance 
\end{itemize}

\section{Data}
\label{subsec:materials}
We use data from the MyPersonality project, which contains the liked Facebook pages of 220,489 volunteers in the United States, along with their scores on the Big 5 personality traits and some personal characteristics such as gender, age, sexual orientation and political preferences~\citep{kosinski2013private}. A Facebook like is a mechanism used by Facebook users to express their positive association with online content, and in this case we focus on the public pages they liked, which can relate to products, public persons, music, sport, books, restaurants, or public statements they agree with. Using this data, it is possible to create a user-like matrix \textbf{$X$} such that $x_{ij} = 1$ if user $i$ liked page $j$.
Behavioral datasets, such as Facebook likes, are usually very sparse as every user can only take a limited number of actions (in this case like Facebook pages), while the total number of possible actions is very large~\citep{junque2013predictive}.
As described in more detail below, we assess the impact of cloaking the likes that lead to the inferences of gender, political orientation and sexual orientation.\footnote{Only gender is still available as an explicit targeting option on Facebook, but machine learning models can still learn the other traits implicitly when optimizing a particular ad or content element.} In this study, we use these as examples of the attributes individuals might wish to safeguard; the specific attributes deemed private of course will vary depending on the individual's preferences.
The data is described in Table~\ref{tab:data}.

\begin{table}[h]
\caption{Data description for the target variables that will be cloaked. We select only the instances that have a value for the correspondent trait.\protect\footnotemark  The features are the Facebook pages that remain after pre-processing. \emph{Active elements} shows the non-zero elements in the entire matrix; \emph{Sparsity} is the percentage of active elements over the total number of elements in the matrix.  \emph{Average likes} is the average number of likes a person associated with this trait has. \emph{Balance} is the percentage of instances with a positive value for the target variable.} \label{tab:data}
\begin{tabular}{@{}l|llllll@{}}
\toprule
Target variable     & Instances & Features & \begin{tabular}[c]{@{}l@{}}Active\\ elements\end{tabular} & \begin{tabular}[c]{@{}l@{}}Sparsity\\ (in \%)\end{tabular} & \begin{tabular}[c]{@{}l@{}}Average\\ likes\end{tabular} & \begin{tabular}[c]{@{}l@{}}Balance\\ (in \%)\end{tabular} \\ \midrule
\textbf{Male} & 165,234    & 115,326  & 16,901,459                                                 & 99.91                                                      & 86.8                                           & 38.37                                                     \\ 
\textbf{Female} & 165,234   & 115,326  & 16,901,459                                                & 99.91                                                     & 112.0                                              & 61.63                                                     \\
\textbf{Homosexual} & 22,477    & 115,326  & 2,197205                                               & 99.92                                                      & 104.3                                                  & 4.67                                                     \\
\textbf{Lesbian} & 29,309    & 115,326 & 4,041,148                                              & 99.88                                                      & 110.5                                                  & 2.65                                                      \\
\textbf{Democrat}   & 36,534   & 115,326 & 4,190,576                                                 & 99.90                                                      & 134.0                                                & 17.27                                                     \\
\textbf{Republican} & 36,534   & 115,326  & 4,190,576                                                & 99.90                                                      & 124.2                                                 & 10.24                                                     \\ 
\bottomrule
\end{tabular}
\end{table}
\footnotetext{For the prediction task of homosexuality, only men whose data record has a value for sexual orientation will be considered, while for the prediction task of lesbian, only women with a value for sexual orientation will be taken into account.}

\paragraph{Personality traits} 
Trait models suggest that personality consists of a range of consistent and relatively stable characteristics (traits) that determine how an individual will think, feel and behave~\citep{matz2016models}. The Big 5 (BF) Model of Personality is the most widely accepted model and proposes five independent traits to capture individual personality differences~\citep{costa1992normal,matz2016models}. The five traits are: 1) \emph{Extraversion}, the tendency to seek excitement and stimulation in the company of others, 2) \emph{Openness}, the tendency to be intellectually curious, creative and unconventional, 3) \emph{Neuroticism}, the tendency to experience negative emotions, and being anxious and nervous 4) \emph{Agreeableness}, the tendency to be trusting, compassionate and cooperative 5) \emph{Consciousness}, the tendency to be organized and efficient~\citep{matz2016models,ramon2021explainable}.
The Big 5 personality traits were established using the international Personality Item Pool (IPIP) questionnaire with 20 items~\citep{goldberg2006international,kosinski2013private}. The traits are recorded on a 5-point Likert scale, and we report the data description in Table~\ref{tab:data_big5}.
Research shows that digital footprints have a predictive power over personality traits which is in line with the typical strength of the relationship between personality and behavior, also known as the \emph{personality coefficient} (a correlation between 0.30 and 0.40)~\citep{meyer2001psychological,azucar2018predicting}.

\begin{table}[h]
\caption{Data description of the Big 5 personality traits. We select only the instances that have a value for the corresponding trait. } \label{tab:data_big5}
\begin{tabular}{@{}l|llllll@{}}
\toprule
Personality trait     & Instances & Features & \begin{tabular}[c]{@{}l@{}}Active\\ elements\end{tabular} & \begin{tabular}[c]{@{}l@{}}Sparsity\\ (in \%)\end{tabular} &  \begin{tabular}[c]{@{}l@{}}Average\\ score\end{tabular} \\ \midrule
\textbf{Extraversion} & 137,529   & 115,326  & 14,513,946                                                & 99.91                                                                                               & 3.55                                                     \\ 
\textbf{Openness} & 137,529    & 115,326  & 14,513,946                                                & 99.91                                                                                                 & 3.88                                                    \\
\textbf{Neuroticism} & 137,529     & 115,326  & 14,513,946                                               & 99.91                                                                                                       & 2.79                                                    \\
\textbf{Agreeableness} & 137,529     & 115,326 & 14,513,946                                              & 99.91                                                                                                       & 3.54                                                      \\
\textbf{Consciousness}   & 137,529    & 115,326 & 14,513,946                                                & 99.91                                                                                                    & 3.42                                                     \\
\bottomrule
\end{tabular}
\end{table}

\section{Cloaking methods}

\paragraph{Cloaking mechanism} As described above, \emph{cloaking} refers to the mechanism of changing user data so that---from the perspective of the inference procedure---it was as if the user did not exhibit this behavior. 
The cloaking mechanism introduced by \citet{chen2017enhancing} relies on counterfactual explanations. These counterfactual explanations explain the decisions made by machine learning models in a human-understandable way, defined specifically as a minimal change to the feature values such that the system's classification decision is changed ~\citep{martens2014explaining,wachter2017counterfactual,verma2020counterfactual,fernandez2022explaining}. When using behavioral data, this corresponds to  a minimal set of active features of the instance, where changing (just) these feature values to zero would lead the model to make a different decision~\citep{ramon2020comparison}.
We apply counterfactual explanations instead of other explanation techniques as they give a direct way to alter the predicted outcome, in line with ~\citet{chen2017enhancing}. Nevertheless, with suitable modifications other explanation techniques such as SHAP could also be adapted to support cloaking~\citep{lundberg2017unified,ramon2020comparison}.
We use the SEDC algorithm\footnote{Python code available at \url{https://github.com/ADMAntwerp/edc}} to compute the counterfactual explanations, which uses a best-first heuristic search strategy to search for the smallest set of features to change~\citep{martens2014explaining,ramon2020comparison}. A change is defined as replacing the original feature value with the median value of that feature over the training data, which in the case of behavioral data will be 0 as this data is usually very sparse. For example, in Facebook data, there is no page that is liked by the majority of users, so the median value of every feature will be 0. Counterfactual explanations will then point to the Facebook likes a user has to cloak (or simply unlike). A user is considered to be successfully cloaked when his or her score falls below the predefined threshold~\citep{chen2017enhancing}.  The average size of a counterfactual explanation, the average number of likes that have to be cloaked to avoid positive inference for each prediction task, can be found in Table~\ref{tab:model_descr}.

\begin{table}[h]
\caption{Model statistics. \emph{Positive rate} indicates the percentage of instances that are predicted as positive by the machine learning model. \emph{AUC} is the model's on the task accuracy, as measured by the area under the ROC curve. \emph{Explanation size} is the average number of likes that must be cloaked to avoid positive inferences.} \label{tab:model_descr} 
\centering
\begin{tabular}{@{}l|ccc@{}}
\toprule
Target variable     & \begin{tabular}[c]{@{}c@{}}AUC\\ (in \%)\end{tabular}   & \begin{tabular}[c]{@{}c@{}}Positive rate\\ (in \%)\end{tabular} & \begin{tabular}[c]{@{}c@{}}Explanation size\\ (avg.)\end{tabular} \\ \midrule
\textbf{Male} & 95.2 & 5.13                                                           & 8                                                                \\
\textbf{Female} & 95.2 & 4.75                                                          & 6                                                                 \\
\textbf{Homosexual} & 89.4 & 5.72                                                           & 4                                                                \\
\textbf{Lesbian} & 77.8 & 7.88                                                          & 2                                                                \\
\textbf{Democrat} & 77.3 & 4.58                                                           & 3                                                                 \\
\textbf{Republican}   & 82.1 & 4.79                                                           & 6                                                                 \\
 \bottomrule
\end{tabular}
\end{table}

\paragraph{Metafeatures} \label{subsec:metafeatures} 
Dimensionality reduction methods reduce the size of the set of features used for modeling~\citep{clark2019unsupervised}.
To group fine-grained features into higher-level metafeatures, we use Non-Negative Matrix Factorization (NMF)~\citep{lee1999learning}. We chose this dimensionality reduction technique as the non-negativity constraint facilitates the interpretation of the extracted metafeatures, and has been shown to provide interpretable results for fine-grained data applications~\citep{contreras2016empirical,ramon2021can}. \footnote{Note that there exist many other techniques to generate the metafeatures, but we do not compare them in this work, as our goal is to study whether using metafeatures can give better performance, rather than to figure out what sort of metafeatures performs best.} We create 50 metafeatures for the Facebook likes and assign each page to the metafeature or topic for which it has the highest weight to ensure mutual exclusivity (each feature only belongs to one metafeature)~\citep{ramon2021can}. An example of two metafeatures is shown in Section~\ref{tab:metafeatures_interpretation}. The creation of these metafeatures is data-driven.  Another option is to use the categories that Facebook assigned to the Facebook like pages itself. These categories are more broad such as `Public Figure' or `Musician/Band'. We will call these domain-based metafeatures, in line with \citet{ramon2021can}. We include the results based on these metafeatures in Appendix~\ref{subsec:results_other_strategies}.

\section{Experimental set-up}

We focus on cloaking the inferences gender (\emph{male} and \emph{female}), sexual orientation (\emph{homosexual} and \emph{lesbian}) and political orientation (\emph{democrat} and \emph{republican}). We train Logistic Regression models with $\ell_2$-regularization with the Scikit-learn library (Python). The literature has shown that this is one of the best performing classification models for behavioral data~\citep{de2020benchmarking}. We use 66\% of the data for training, and the remaining 33\% for testing. We also exclude users with fewer than 10 likes and Facebook pages with fewer than 10 likes. For fine-tuning the hyperparameters of the model, we perform a grid search on the training set by using three-fold cross-validation, where we tune the regularization parameter C of $\ell_2$-LR model. As is common in targeted advertising, we assume that a positive inference is drawn, which means that the user would be subjected to targeted advertising, when the model assigns the user a score which places him or her in a specified top quantile of the score distribution produced by the prediction model~\citep{chen2017enhancing,perlich2014machine}. For online targeting, a typical value for this quantile is between 90 and 100, and we base our threshold for positive inference on the top 95 quantile in the training set~\citep{chen2017enhancing, perlich2014machine}.  The  AUC and positive rate for each prediction task can be found in Table~\ref{tab:model_descr}. 

\subsection{Longer-term cloaking protection}

\begin{figure}[h]
    \centering
    \includegraphics[width=0.90\textwidth]{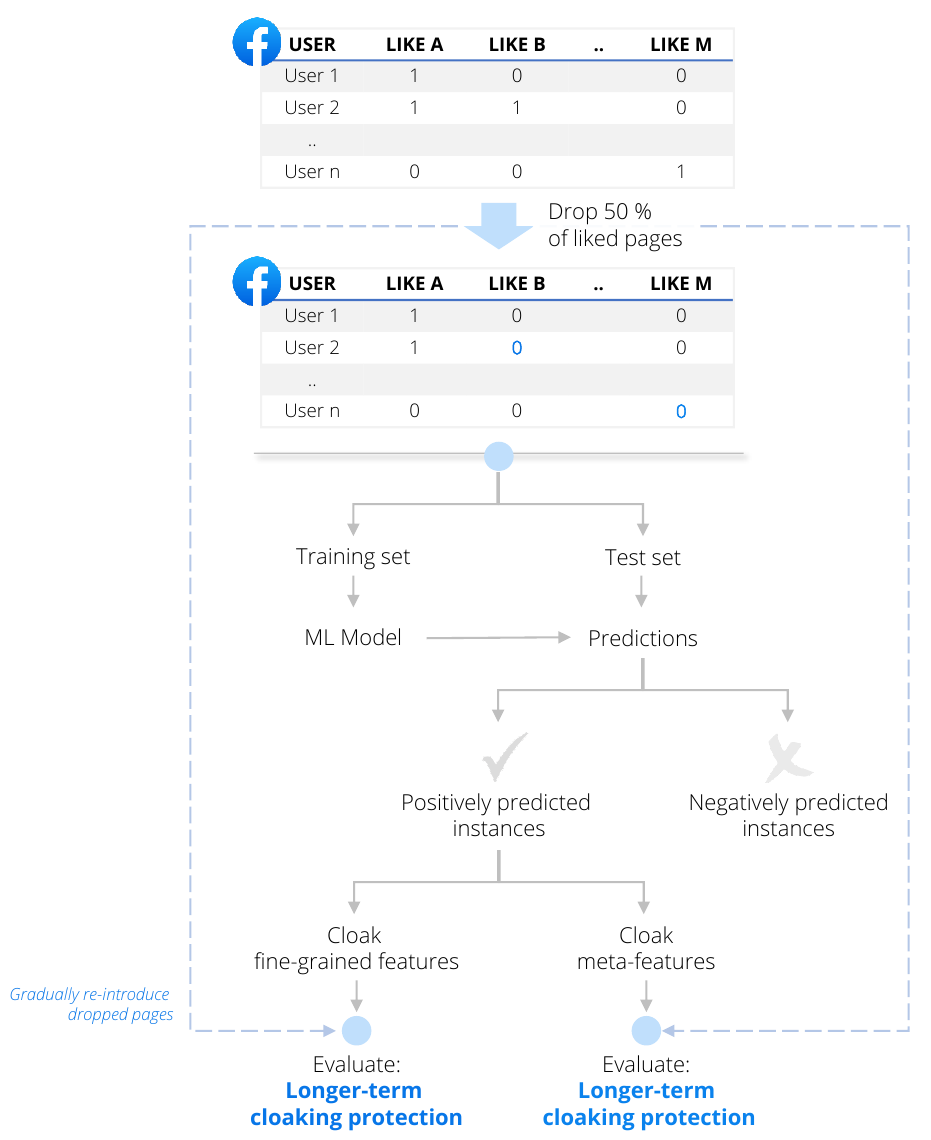}
    \caption{Experimental set-up to measure the longer-term cloaking protection.}
    \label{method_exp1}
\end{figure}

We study longer-term cloaking protection using the methodology described in Figure~\ref{method_exp1}. We simulate a person's behavior over time by first dropping 50\% of likes for each user at random.\footnote{This means that the simulation uses the assumption that people's behavior over time is stable in the short run, as their liking behavior does not change---since the data does not include time stamps. Verifying this on time-stamped data would be an avenue for follow-up research.}
After dropping the pages, we train a regularized logistic regression model for every prediction task on the reduced training set. We use this model to make predictions on the instances in the reduced test set and select the positively predicted instances.  For these instances, we compare two cloaking strategies to remove positive inference:
\begin{enumerate}
    \item Cloaking the individual likes, which we also call the fine-grained features (where we remove all the pages in the counterfactual explanation of that instance as described by \citet{chen2017enhancing}). We call this strategy \emph{FG}.
    \item Cloaking the metafeatures, where we remove all the pages in the counterfactual explanation of that instance \textbf{and} the pages that belong to the same metafeatures as the pages in the counterfactual explanation. We call this strategy {MF}.
\end{enumerate}
Using the second strategy also means that the number of liked pages will decrease substantially more than when using the first strategy, which we will show in Figure~\ref{fig:trade-off}.
We simulate an individual's behavior over time by gradually re-introducing more of the 50\% of pages that initially were dropped. We measure the \textbf{longer-term cloaking protection} of both strategies by measuring the percentage of positively predicted instances for which cloaking this targeting task in the past successfully inhibited future inferences for that same task and individual.

\subsection{Trade-off between privacy and personalization} \label{subsec:set_up_exp2}
\begin{figure}[h]
    \centering
    \includegraphics[width=0.80\textwidth]{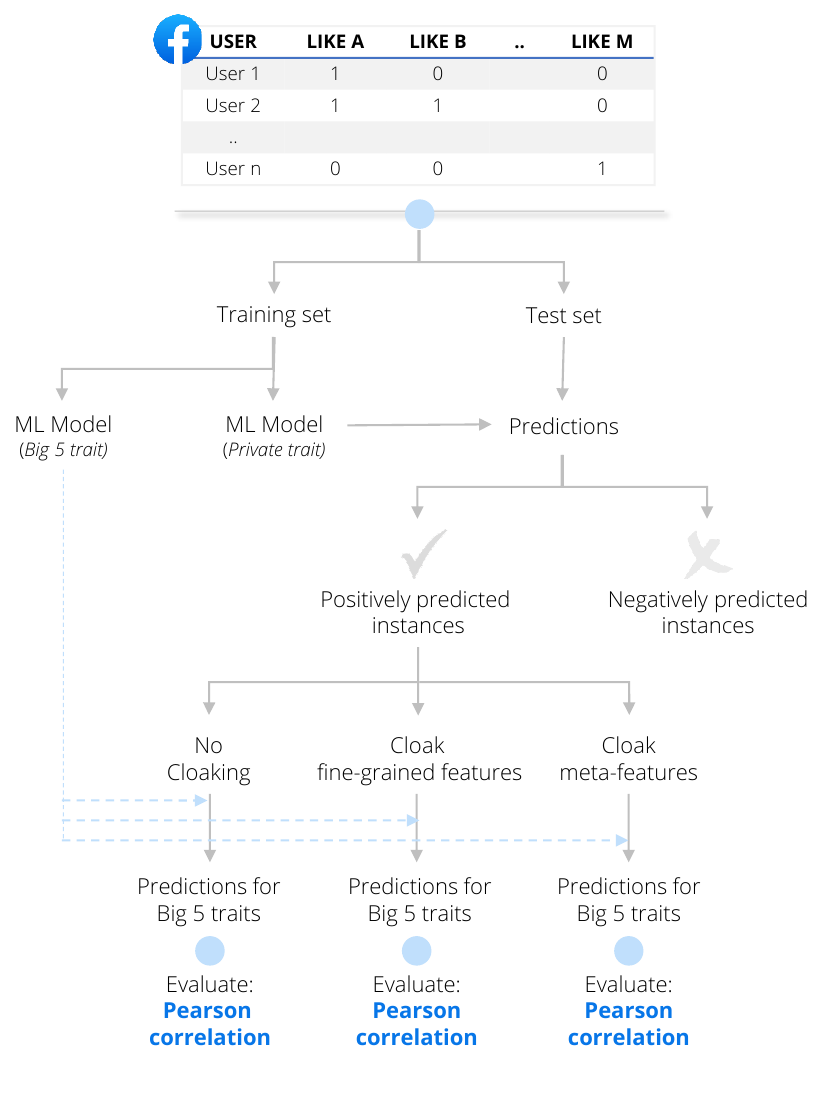}
    \caption{Experimental set-up to measure the impact of cloaking a private trait on other prediction tasks. }
    \label{method_exp2}
\end{figure}

As introduced above, although we may think myopically about a privacy-preserving action when taking it, such actions can have broader effects.  In particular, cloaking data in order to inhibit one inference can have effect on other inferences---possibly ones that we would not want to inhibit.  Therefore, consumers and platforms should be interested in what the effect of hiding larger parts of someone's traces is on the performance of other prediction tasks. 

Define $X$ as the initial complete data, and $X_c$ as the cloaked data. To what extent does changing $X$ to $X_c$ affect the predictions of models for different (but desired) target variables?

We describe the set-up of our experiment in Figure~\ref{method_exp2}.  We examine the effect on a second set of prediction tasks when cloaking for the trait-prediction tasks we described above.  The new tasks involve predicting an individual's ratings for the Big 5 personality traits, the accuracy of which we measure using Pearson correlation, which is the most commonly used measure of prediction accuracy for predicting these personality traits~\citep{kosinski2013private,azucar2018predicting}.  We choose the Big 5 traits as the second set of tasks because they cover broad aspects of personality and are very well understood. We compare the effect of not cloaking an individual's data, cloaking fine-grained footprints (FG), and cloaking metafeatures (MF).\footnote{The point here is not that inferences for the Big 5 traits are not privacy invasive; this of course will depend on the individual.  Rather, the point simply is to examine the effects of cloaking some potentially sensitive inferences on other inferences that are broadly applicable.}


\section{Results: Longer-term cloaking protection}
\begin{figure}[h]
    \centering
    \includegraphics[width=0.99\textwidth]{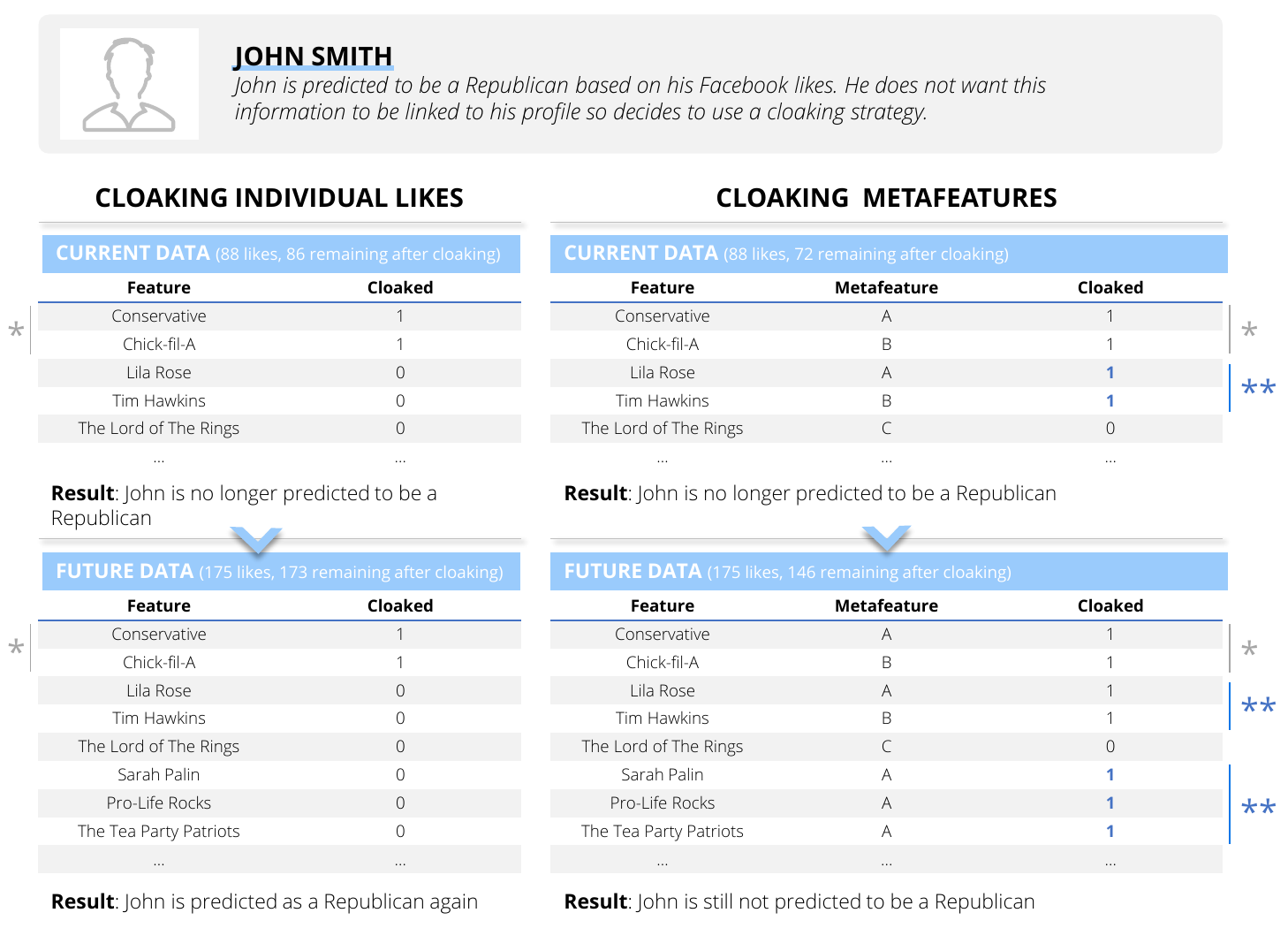}
    \caption{Example of John. The column \emph{Cloaked} signals the pages that are cloaked for each strategy and point in time.\\
    *: Original features cloaked to ensure John is not predicted as republican.\\
    **: Additional features cloaked because part of same metafeature as the original features.}
    \label{fig:john}
\end{figure}

Imagine a random person, John, who has been using a particular technology platform and thereby leaving digital footprints.  The platform's political-orientation model gives him a high enough score as a \emph{republican} in order for him to receive corresponding political ads. John no longer wants to receive advertisements related to his political orientation; maybe because he no longer identifies as such, or he wants to keep his political orientation private, or he simply finds the advertisements annoying.
We want to see if, as he subsequently continues to like pages, John gets targeted as republican again after using the cloaking device described by \citet{chen2017enhancing}.


We represent John in Figure~\ref{fig:john}. As described above, we simulate the point where the model uses half of his digital traces (88 likes) as the current footprint data, and the point with all his digital footprints as his future data (175 likes). Using the current data, John is predicted as a republican.\footnote[2]{Prediction score = 0.161, which is above the targeting threshold of 0.148.}
John wants to inhibit this inference and receives the following advice (based on counterfactual explanations to bring him under the threshold):
\emph{If you would hide the likes `Conservative' and `Chick-fil-A`, you would no longer be predicted as republican.}
After cloaking these pages, John has 86 likes remaining and is no longer predicted as republican.~\footnote[3]{Prediction score = 0.140, which is below the threshold of 0.148.}
We move on to the future point. John, who has remained active, has liked 87 new pages (essentially doubling his digital traces). Even though the two liked pages from his initial counterfactual explanation are still cloaked, John gets re-predicted as republican due to his new digital footprints.~\footnote[8]{His prediction score on the full data is 0.260; after cloaking the two likes in his explanation, his prediction score is 0.225.  This above the targeting threshold of 0.194. The threshold is different now because after everyone in the dataset has acquired new digital traces, the scores for the top 5th percentile will be different.}  This illustrates that cloaking is not necessarily robust in the longer term, as individuals continue to leave new digital footprints.  Note that \citet{chen2017enhancing} pointed out as a limitation of the cloaking design that if cloaking does not also cover closely associated features, one might end up being targeted again in the future~\citep{chen2017enhancing}.

We introduced cloaking based on metafeatures in an attempt to (partially) address this lack of robustness.  Recall that cloaking metafeatures also cloaks other footprints that are (estimated to be) closely related to those suggested by the counterfactual explanation.  So for our current example, the Facebook page `Conservative' belongs to metafeature A, and the Facebook page `Chick-fil-A' belongs to metafeature B. Typically, metafeatures such as these\footnote{Specifically, those created by embedding the original data in a lower dimensional space.} are interpreted by looking at the top weighted fine-grained features for each metafeature~\citep{wang2012nonnegative,o2015analysis,contreras2016empirical}. These are shown in Table~\ref{tab:metafeatures_interpretation}. 
\begin{table}[h]
\caption{Interpretation of two metafeatures generated with NMF by showing the 10 features with the highest coefficients for each metafeature.} \label{tab:metafeatures_interpretation}
\begin{tabular}{@{}|l|l|@{}}
\toprule
\textbf{Metafeature A}      & \textbf{Metafeature B}                               \\ \midrule
\textit{Being Conservative} & \textit{The Bible}                                   \\
\textit{Sarah Palin}        & \textit{Jesus Daily}                                 \\
\textit{Conservative}       & \textit{``I'm proud to be Christian" by Aaron Chavez} \\
\textit{Glenn Beck}         & \textit{Casting Crowns}                              \\
\textit{Fox News}           & \textit{Chris Tomlin}                                \\
\textit{Tea Party Patriots}           & \textit{Third Day}                                \\
\textit{Mitt Romney}           & \textit{TobyMac}                                \\
\textit{FreedomWorks}           & \textit{Jeremy Camp}                                \\
\textit{Sean Hannity}           & \textit{Switchfoot}                                \\
\textit{John McCain}           & \textit{Skillet Music}                                \\ \bottomrule
\end{tabular}
\end{table}

Metafeature A clearly is related to right-wing politics, and metafeature B to Christianity. Metafeature cloaking hides not only those likes (fine-grained features) in the counterfactual explanation, but also all the likes that belong to the same metafeature as each of these likes. 
When we also cloak all the likes in the associated metafeatures, 14 additional pages are cloaked.  These include `Tim Hawkins' and `Lila Rose'.\footnote{This brings the prediction score further down to 0.120.} Subsequently, when John likes pages in the future, the new pages associated with those same metafeatures will be hidden as well. For John, this leads to also hiding pages such as `Sarah Palin', `Tea Party Patriots' and `Pro-Life Rocks'. In total, 29 new pages are cloaked in the future and the result is that John will not be predicted as republican even after leaving his future footprints.\footnote{The prediction score on the future data after cloaking the metafeatures is 0.126, which is well below the threshold of 0.194.}

\begin{figure}[!htb]
    \centering
    \begin{subfigure}[b]{0.47\textwidth}
         \centering
         \includegraphics[width=\textwidth]{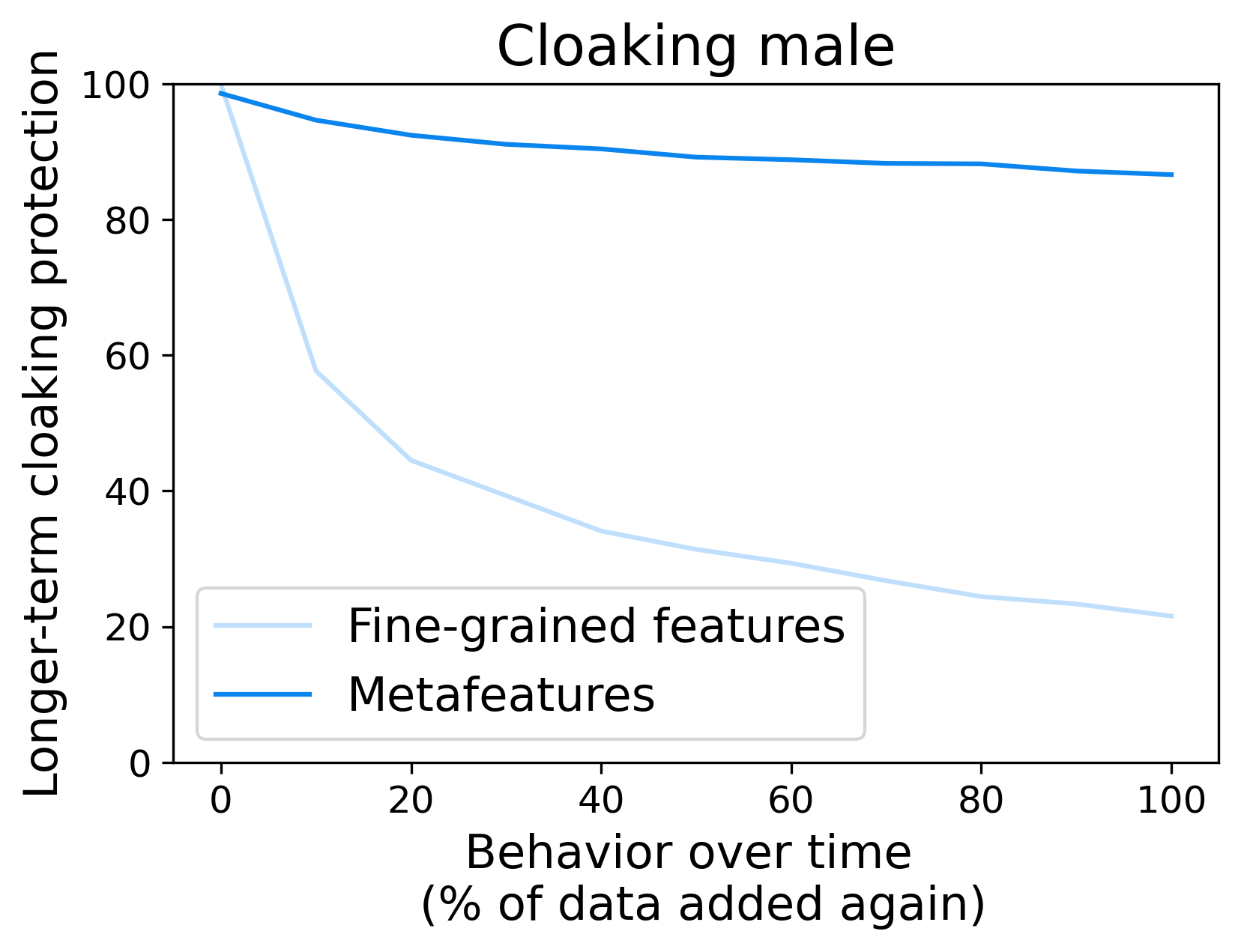}
         \label{fig:y equals x}
     \end{subfigure}
     \hfill
     \begin{subfigure}[b]{0.47\textwidth}
         \centering
         \includegraphics[width=\textwidth]{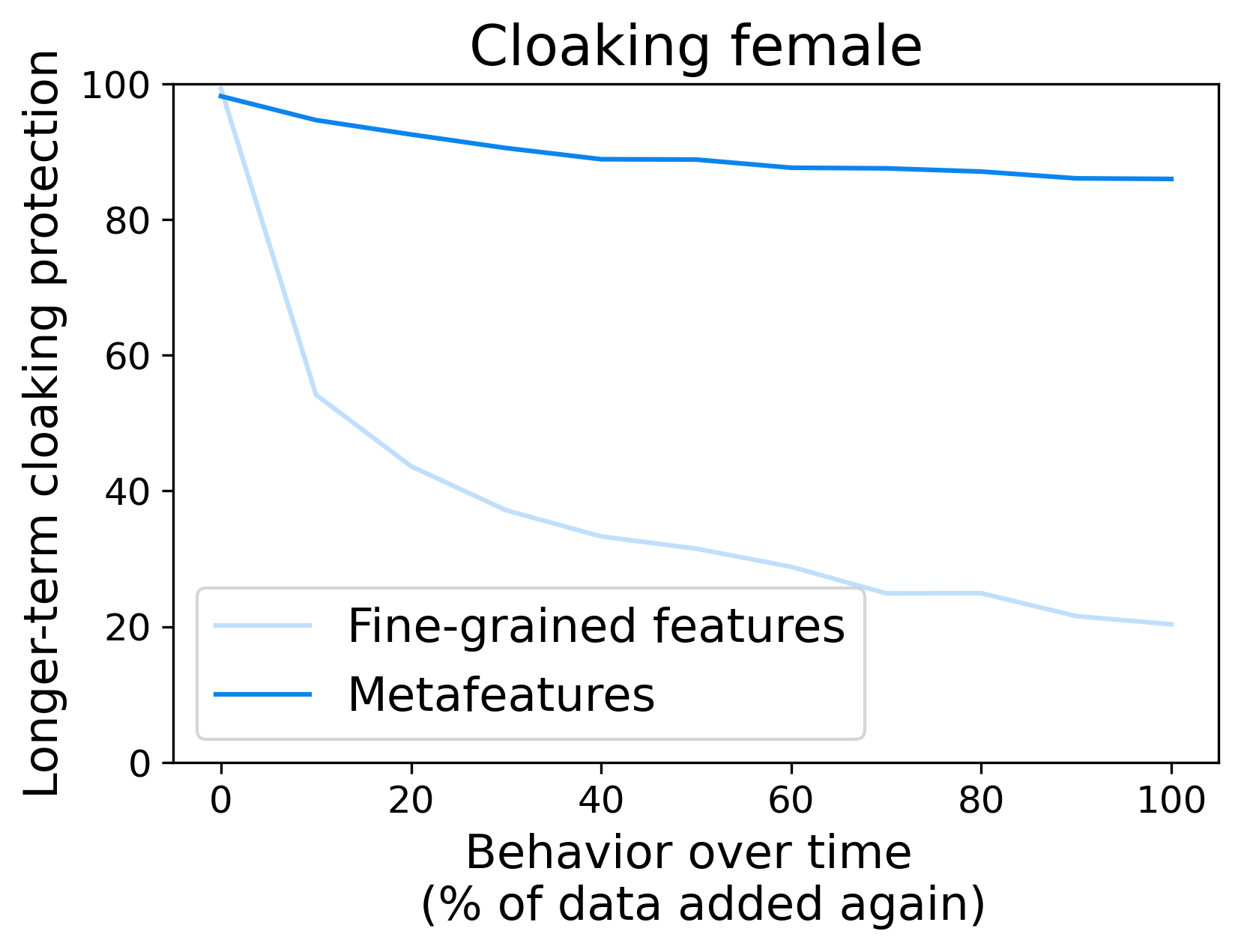}
         \label{fig:y equals x}
     \end{subfigure}
     \hfill
     \begin{subfigure}[b]{0.47\textwidth}
         \centering
         \includegraphics[width=\textwidth]{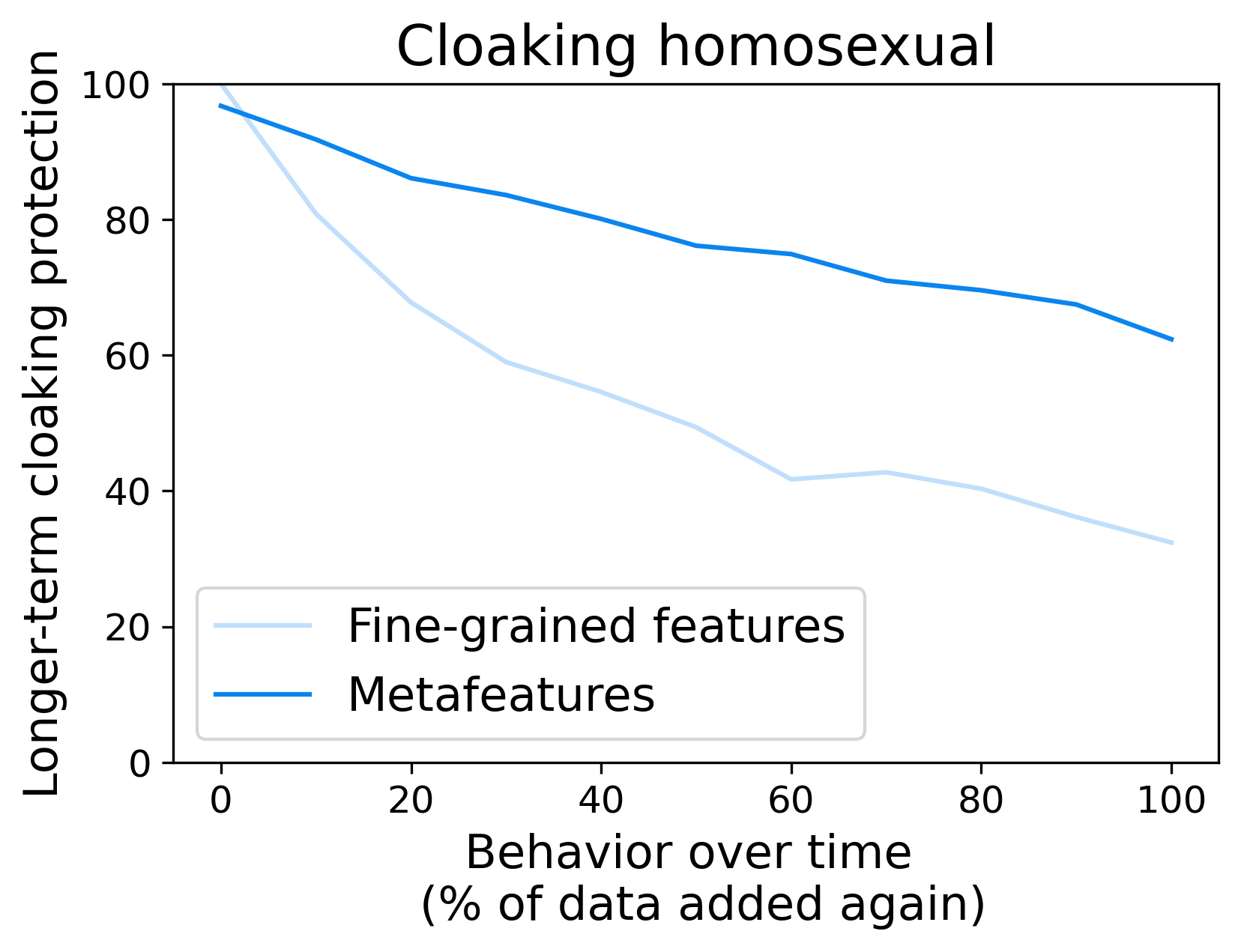}
         \label{fig:y equals x}
     \end{subfigure}
     \hfill
     \begin{subfigure}[b]{0.47\textwidth}
         \centering
         \includegraphics[width=\textwidth]{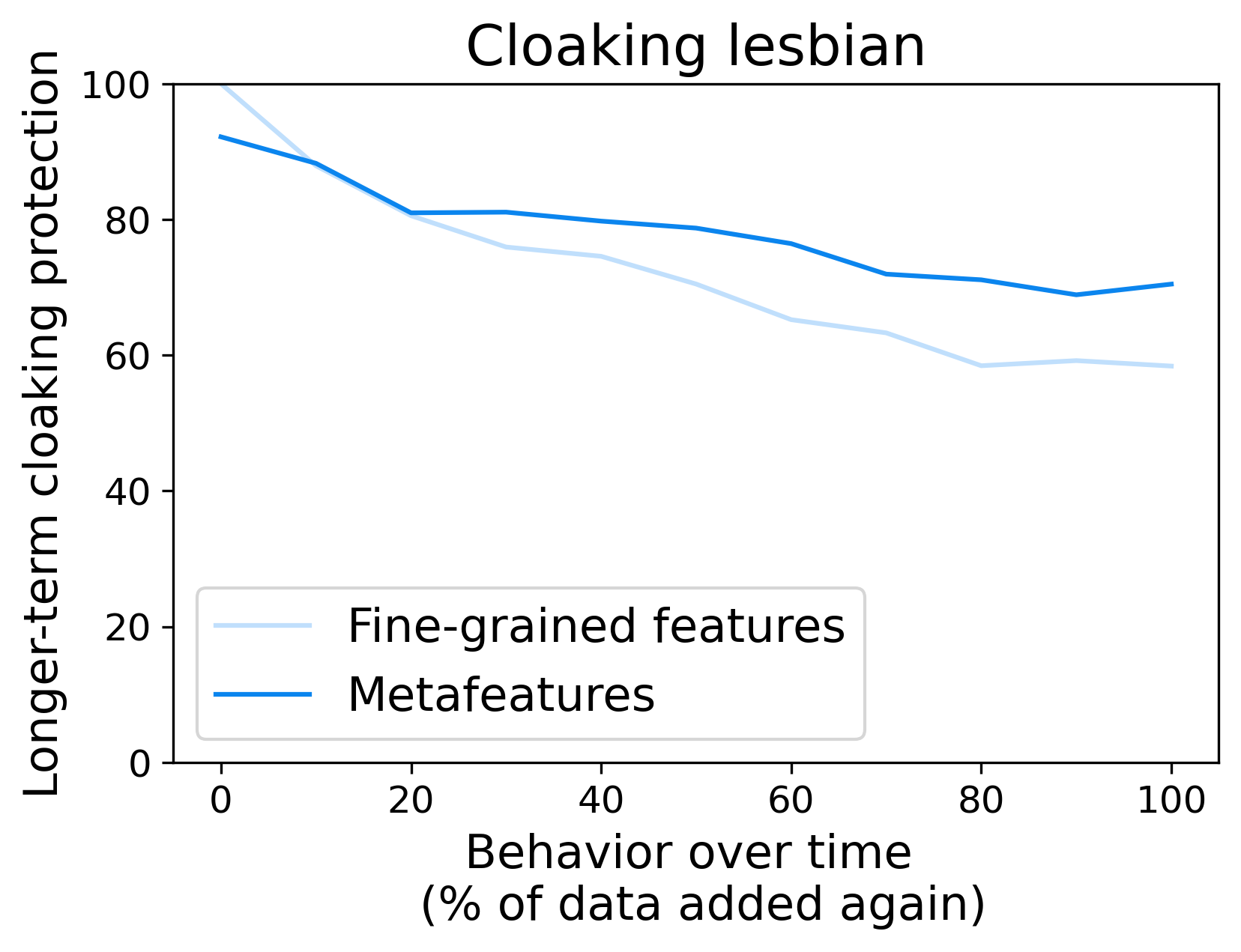}
         \label{fig:y equals x}
     \end{subfigure}
     \hfill
     \begin{subfigure}[b]{0.47\textwidth}
         \centering
         \includegraphics[width=\textwidth]{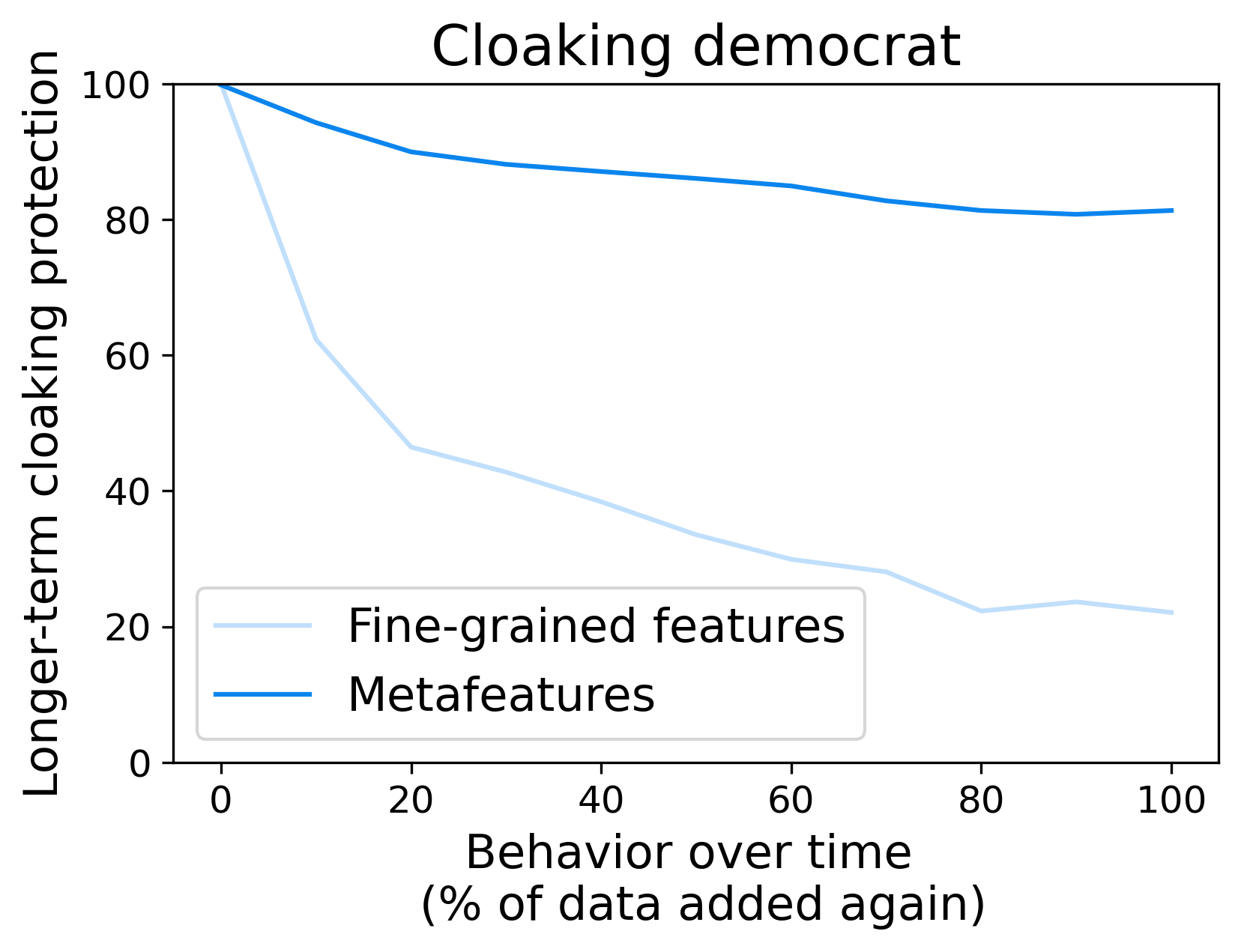}
         \label{fig:y equals x}
     \end{subfigure}
     \hfill
     \begin{subfigure}[b]{0.47\textwidth}
         \centering
         \includegraphics[width=\textwidth]{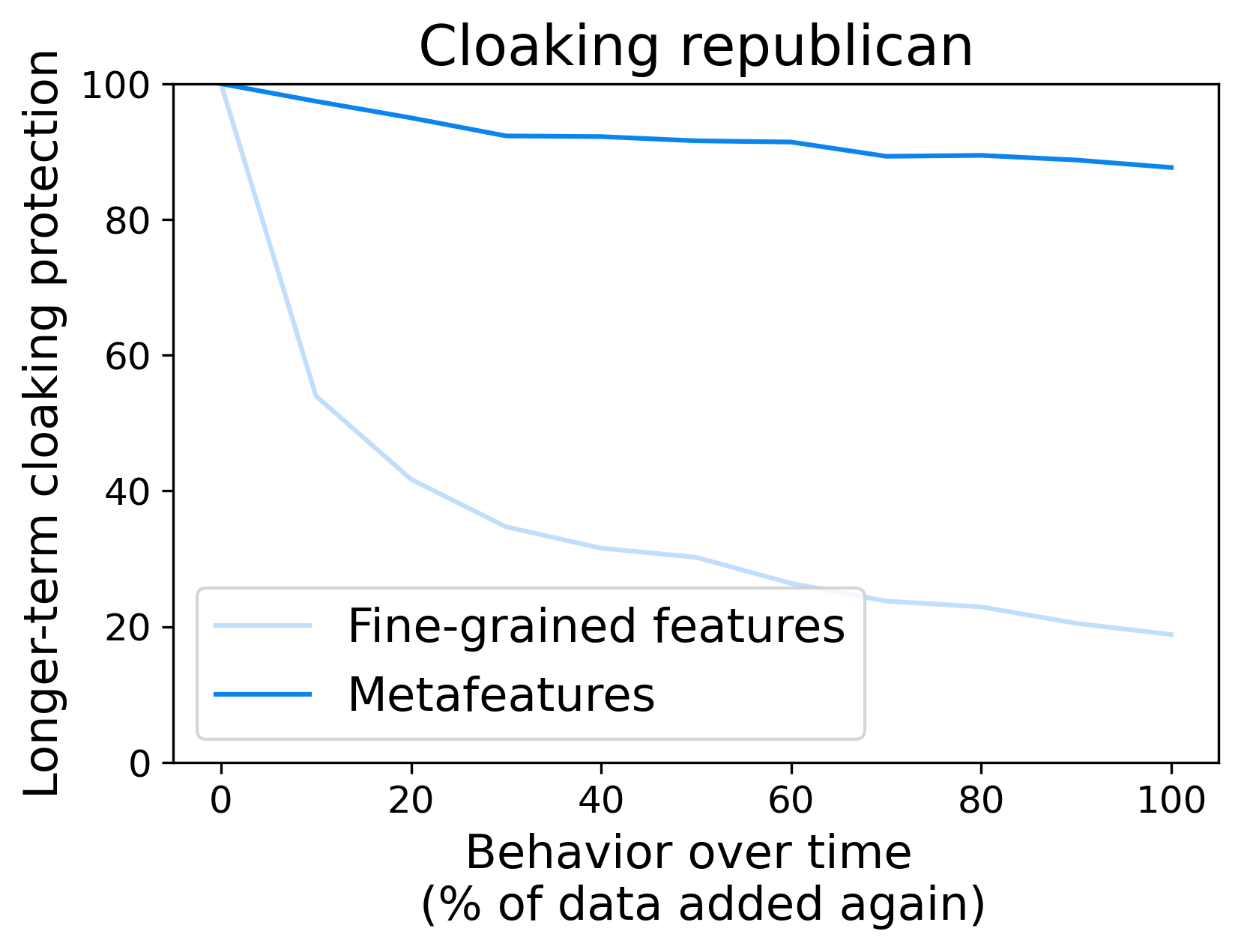}
         \label{fig:y equals x}
     \end{subfigure}
     \hfill
    \caption{Longer-term cloaking protection. We measure the longer-term cloaking protection as the percentage of positively predicted instances for which cloaking this targeting task successfully inhibits future inference. The population taken into account constitutes the intersection of individuals predicted as positive when using 1/2 of the data, and when using the full data.
     We measure the evolution over time on the x-axis by gradually re-adding the dropped pages.} 
    \label{fig:exp1_results}
\end{figure}

Moving beyond the specific example of John, we compare the longer-term cloaking protection of the two cloaking strategies in hiding gender, political orientation and sexual orientation.
As shown in Figure~\ref{fig:exp1_results}, cloaking the fine-grained features offers less protection over time than cloaking the metafeatures. People get targeted again relatively quickly. For example, when cloaking \emph{male}, after adding 10\% new likes, only 57.6\% of instances are still successfully cloaked. After adding all their new likes (and thus doubling their digital traces), only 21.5\% are still successfully cloaked. On the other hand, when we cloak the metafeatures instead, we see that 86.6\% are still successfully cloaked when the digital traces are doubled.

We see the same patterns when cloaking \emph{female} and political orientation (\emph{democrat} and \emph{republican}).
Sexual orientation, especially \emph{lesbian}, is more effectively cloaked over time than other tasks when using fine-grained features; cloaking the metafeatures is still a more effective longer-term cloaking strategy, but the difference between the strategies is be smaller.\footnote{We see that for the prediction task of \emph{lesbian}, for a very small number of individuals, cloaking the metafeatures instead of the fine-grained can lower the number of successfully cloaked individuals, even without adding additional data.} 
We hypothesize that this could be 
because there are fewer people whose true target label is \emph{lesbian} in the targeted population (\emph{True Positives}).

We analyze whether there is a difference in longer-term cloaking protection between correctly predicted people (True Positives) and people that were incorrectly predicted as the target variable (False Positives). Are people who were correctly inferred more likely to reveal themselves again over time?
\begin{figure}[ht]
    \centering
     \begin{subfigure}[b]{0.47\textwidth}
         \centering
         \includegraphics[width=\textwidth]{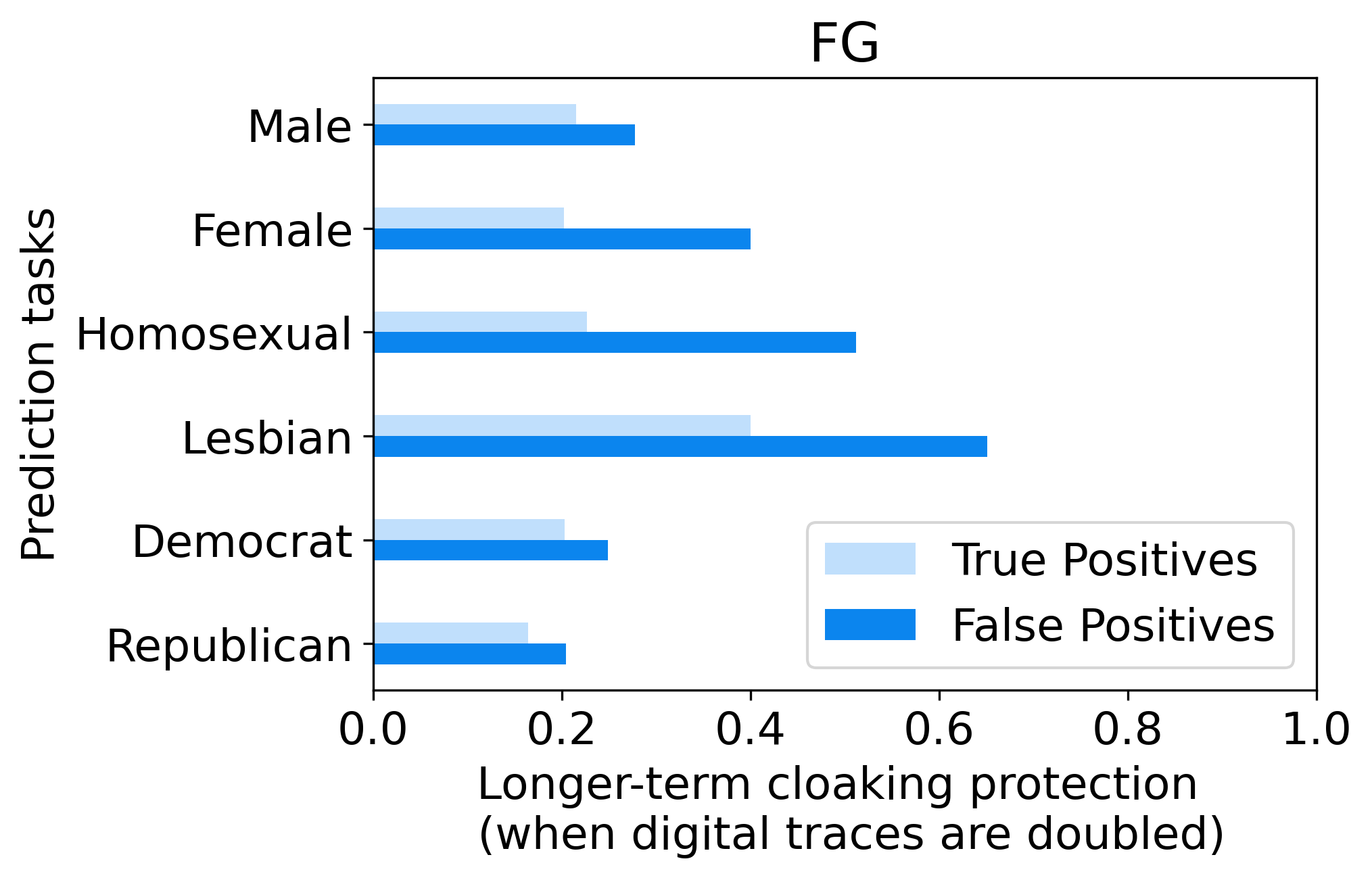}
         \caption{}
         \label{fig:tp_fg}
     \end{subfigure}
     \hfill
     \begin{subfigure}[b]{0.47\textwidth}
         \centering
         \includegraphics[width=\textwidth]{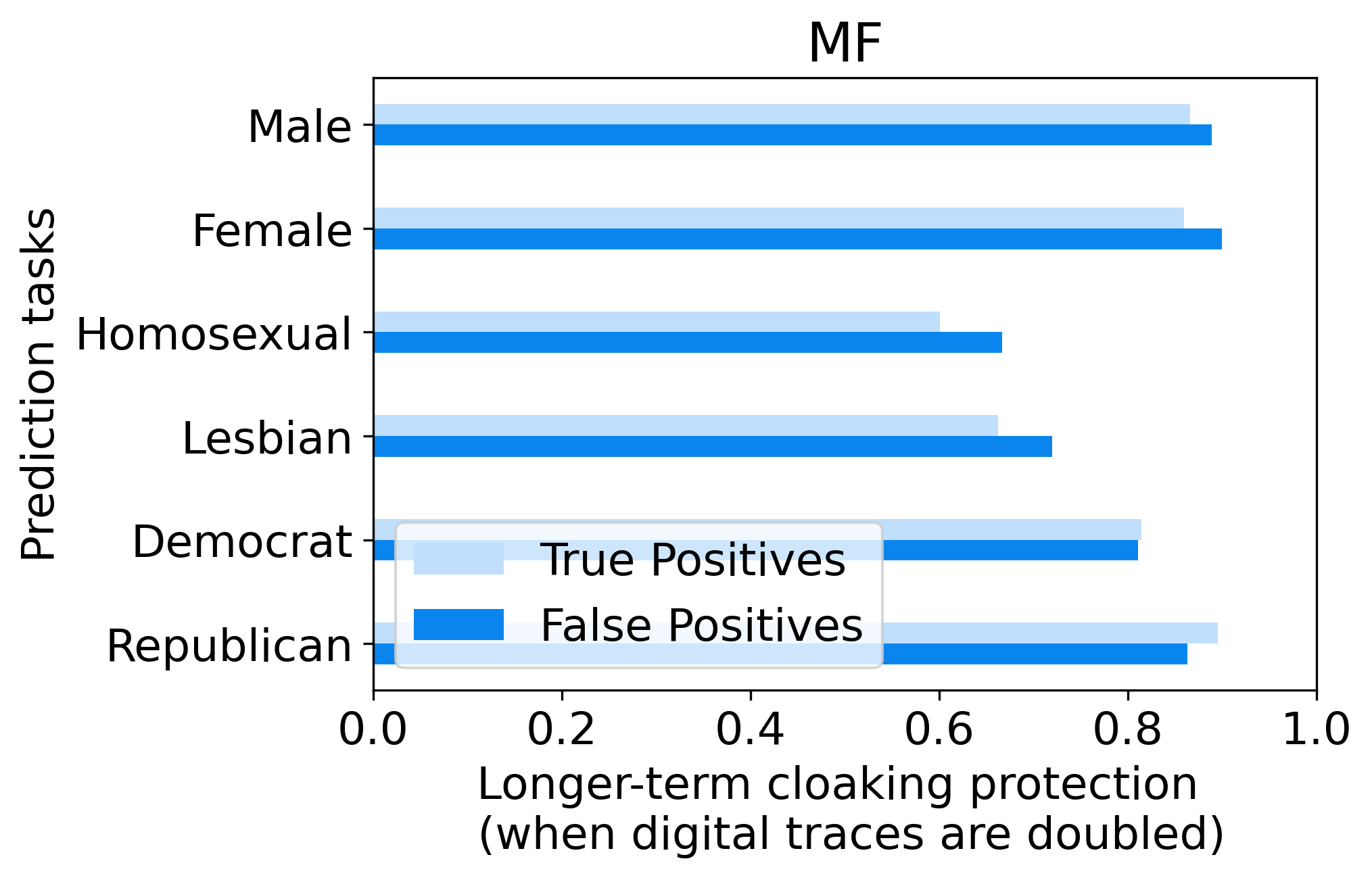}
         \caption{}
         \label{fig:tp_mf}
     \end{subfigure}
     \hfill
    \caption{Is there a difference in longer-term cloaking protection between True Positives and False Positives? We measure this at the point where all the initially dropped data is re-added and the digital traces are thus doubled.} \label{fig:exp1_tp}
\end{figure}
In Figure~\ref{fig:tp_fg}, one can observe that the longer-term cloaking protection of True Positives is in fact lower. This aligns with intuition, given their higher likelihood of repeating behaviors that could result in the same prediction. This difference almost disappears when we assess the longer-term cloaking protection with metafeatures (Figure~\ref{fig:tp_mf}).

We also present the results of two additional cloaking strategies in Appendix~\ref{subsec:results_other_strategies}. The first option involves using the categories assigned by Facebook to the like pages themselves, which we refer to as domain-based metafeatures. The advantage of using domain-based metafeatures is that they are readily available and by design comprehensible. However, as shown in Figure~\ref{fig:exp1_domain}, the data-driven metafeatures created by NMF are more effective in avoiding inferences over time, and in addition our analysis reveals that on average they hide fewer pages than the domain-based metafeatures. We conjecture that this is because they more accurately capture general patterns of behavior. For example, when examining the metafeatures in Table~\ref{tab:metafeatures_interpretation}, we see that they are strongly associated with right-wing politics and Christianity, which are both highly predictive of being a Republican. On the other hand, domain-based metafeatures such as `Public Figure' may be too general to capture these specific patterns. 

Another cloaking strategy we experiment with involves adding a \emph{tolerance level} to the initial counterfactual explanations. This means that instead of using the threshold of the decision-making system to generate the counterfactual explanations for cloaking, we employ a \textbf{lower} threshold. It is to be expected that when we bring someone just below the threshold (which is what counterfactual explanations do), the chances of them crossing the threshold again are relatively high. Therefore, we explore bringing individuals not only below the 95\% threshold but also below the 90\% quantile (while still using the 95\% quantile as threshold for predictions). This approach should provide  an additional layer of protection from future targeting. As depicted in Figure~\ref{fig:exp1_tolerance}, it does indeed offer extended protection initially, but on average, the protection of the cloaking strategy still diminishes rapidly as more likes are accumulated over time.
This strategy has no impact on the pages that will be liked in the future, and this is clearly evident in the results. This highlights one of the major advantages of using metafeatures as a cloaking strategy.

\section{Results: Trade-off between privacy and personalization} \label{subsec:cross-target}

\begin{figure}[!htb]
    \centering
    \includegraphics[width=0.7\textwidth]{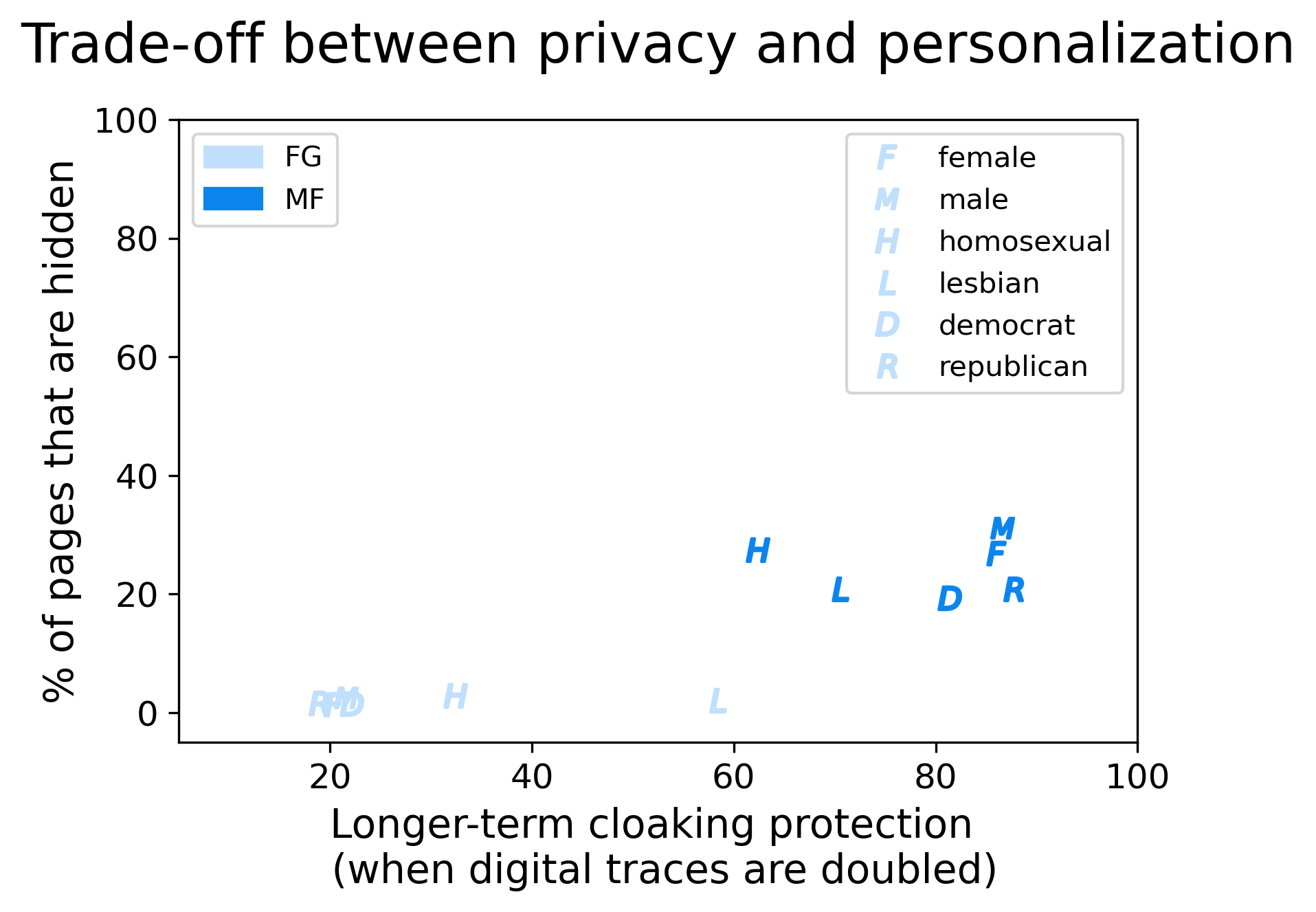}
    \caption{What percentage of people's digital footprints are hidden with each cloaking strategy? We measure loss in personalization by the average \% of someone's likes that have to be removed, and privacy protection as the level of longer-term cloaking protection at the point when the individual's digital footprints are doubled.}
    \label{fig:trade-off}
\end{figure}
Cloaking metafeatures hides larger portions of an individual's digital footprints, resulting in increased privacy protection but potentially losing the benefits of personalization. We assume users do not want to lose all personalization; otherwise, an individual could simply cloak all his liked pages and no inferences would be made (this could still be a viable option for some users, although this will be a bad outcome from the perspective of the advertising platform). 
Figure~\ref{fig:trade-off} illustrates that cloaking metafeatures results in a substantial increase in privacy, but also in a substantial increase in the number of pages that are being hidden than when using fine-grained features. In the example of John, after cloaking the fine-grained features, he has 173 likes left for personalization, while after cloaking the metafeatures, he only has 149 likes left. Therefore it is important to assess the impact of cloaking an individual's sensitive traits on the ability to predict other things about the individual.

\begin{figure}[!htb]
    \centering
    \begin{subfigure}[b]{0.47\textwidth}
         \centering
         \includegraphics[width=\textwidth]{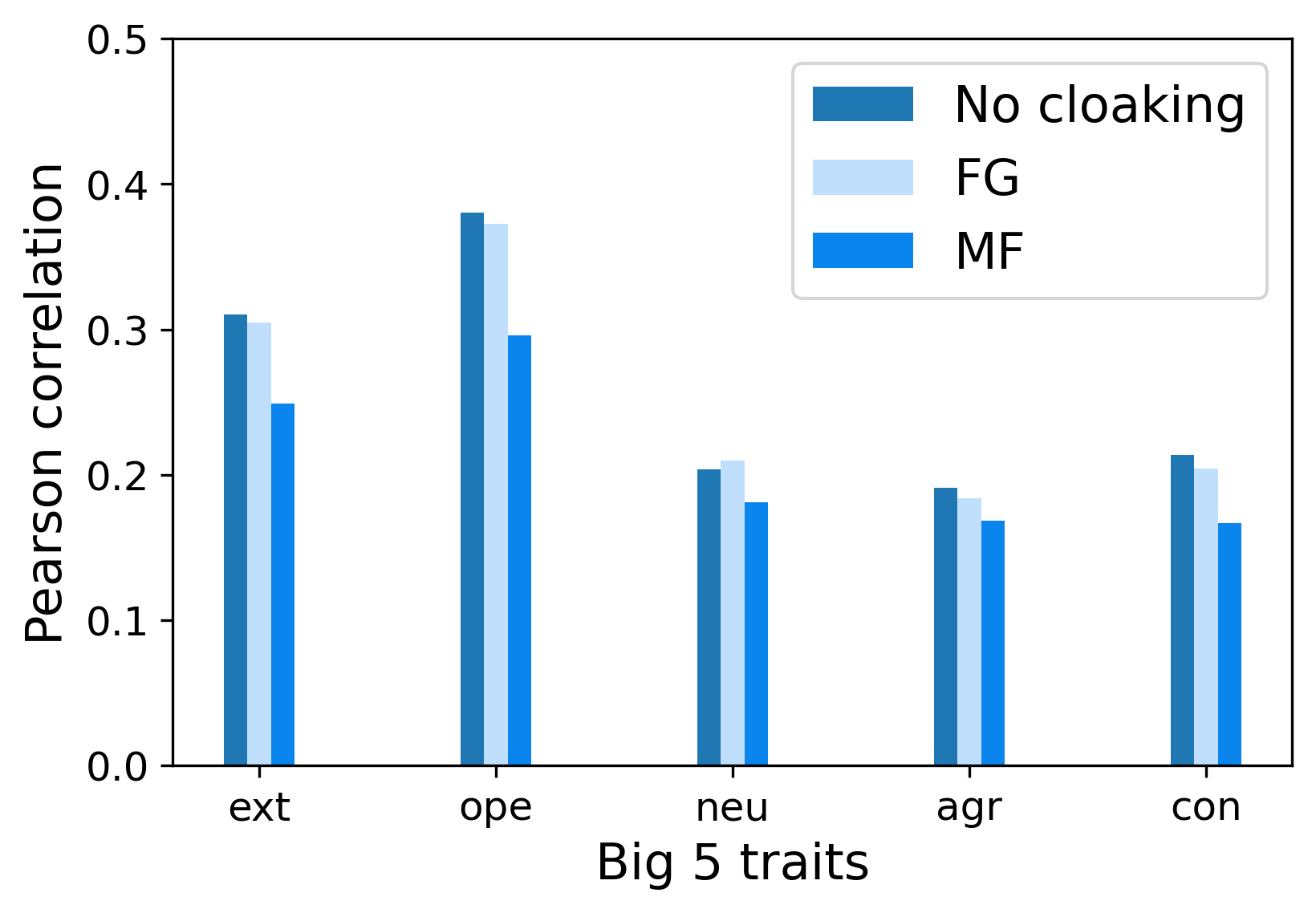}
         \caption{Male }
         \label{fig:y equals x}
     \end{subfigure}
     \hfill
     \begin{subfigure}[b]{0.47\textwidth}
         \centering
         \includegraphics[width=\textwidth]{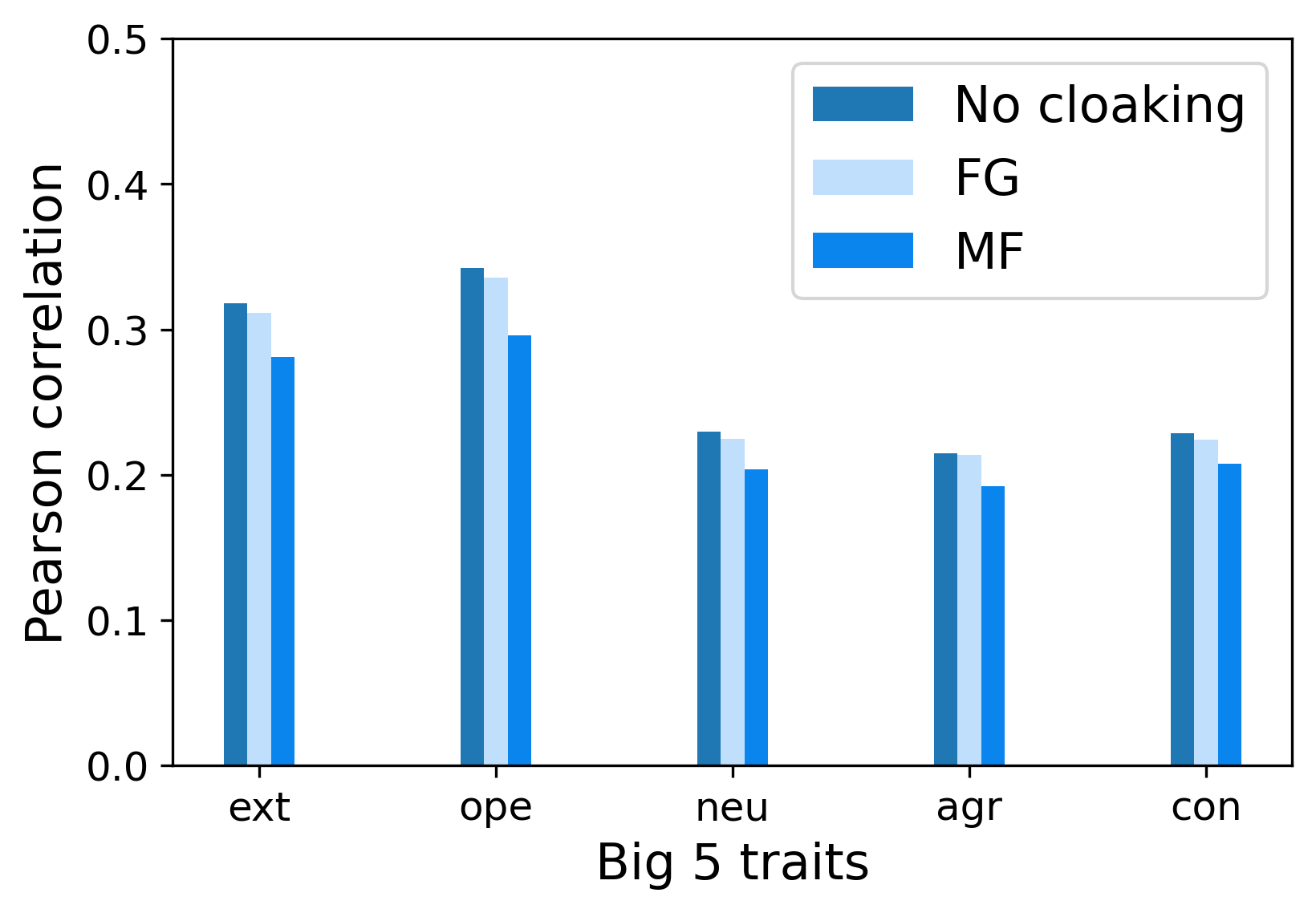}
         \caption{Female }
         \label{fig:y equals x}
     \end{subfigure}
     \hfill
     \begin{subfigure}[b]{0.47\textwidth}
         \centering
         \includegraphics[width=\textwidth]{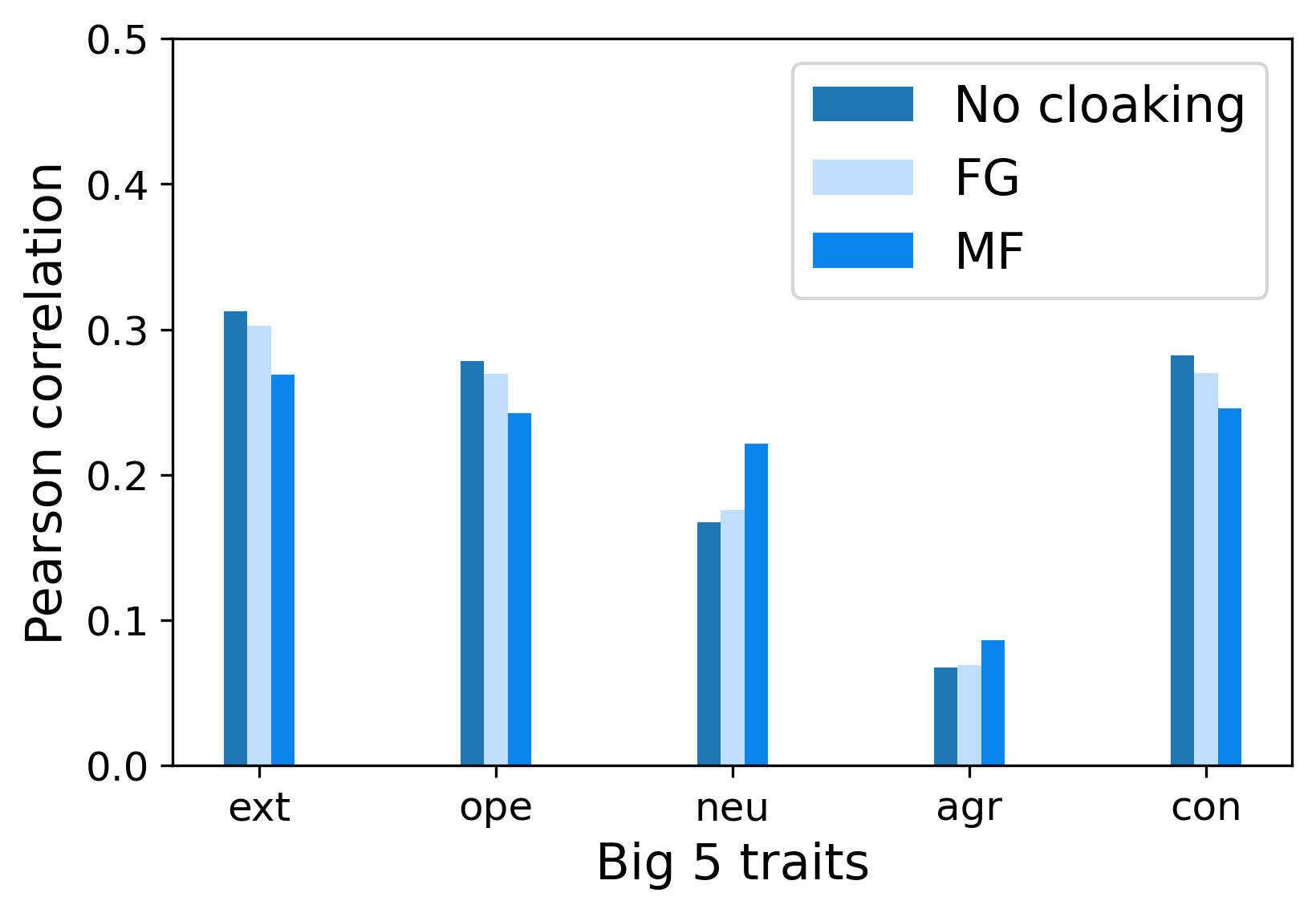}
         \caption{Homosexual}
         \label{fig:y equals x}
     \end{subfigure}
     \hfill
     \begin{subfigure}[b]{0.47\textwidth}
         \centering
         \includegraphics[width=\textwidth]{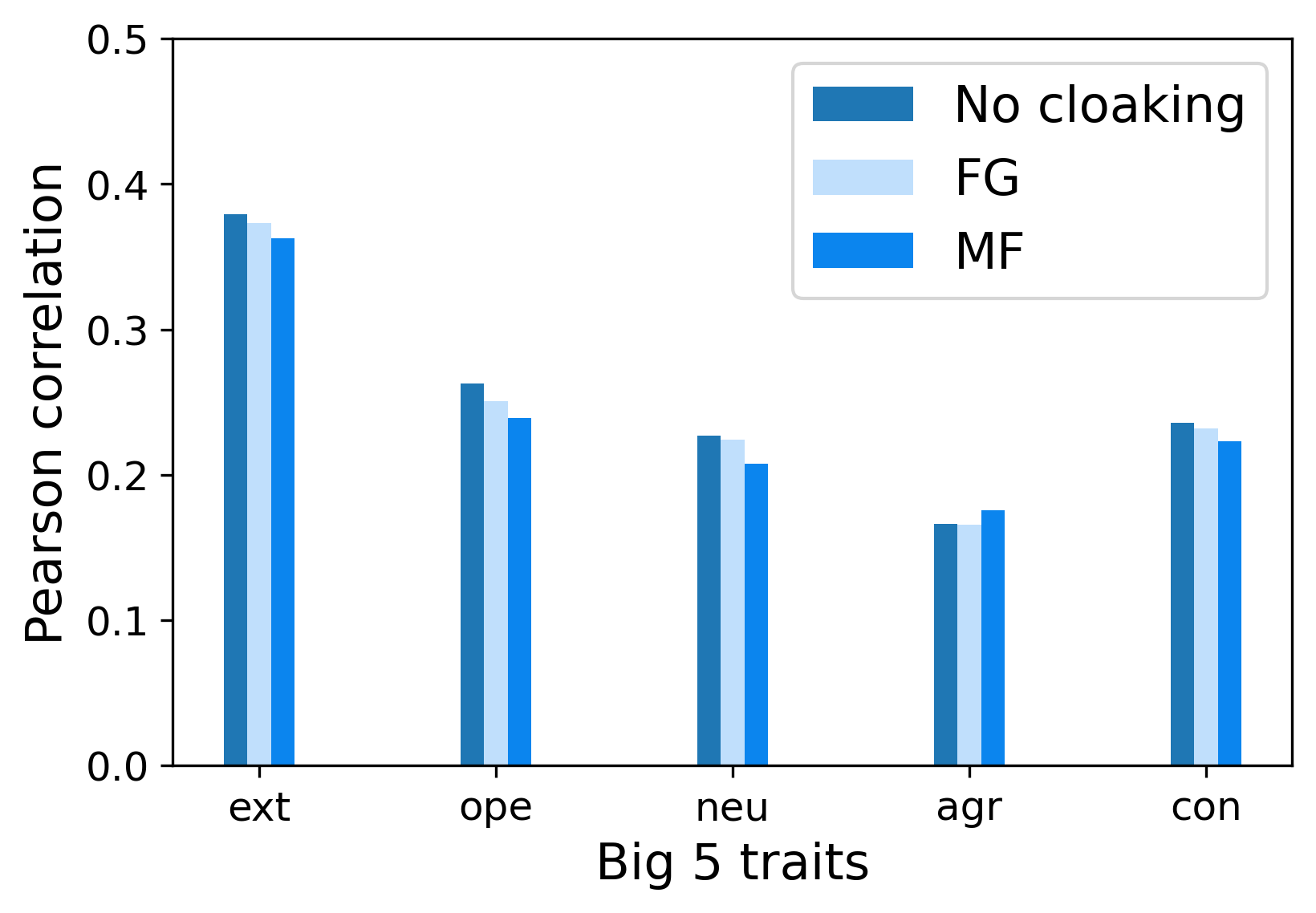}
         \caption{Lesbian}
         \label{fig:y equals x}
     \end{subfigure}
     \hfill
     \begin{subfigure}[b]{0.47\textwidth}
         \centering
         \includegraphics[width=\textwidth]{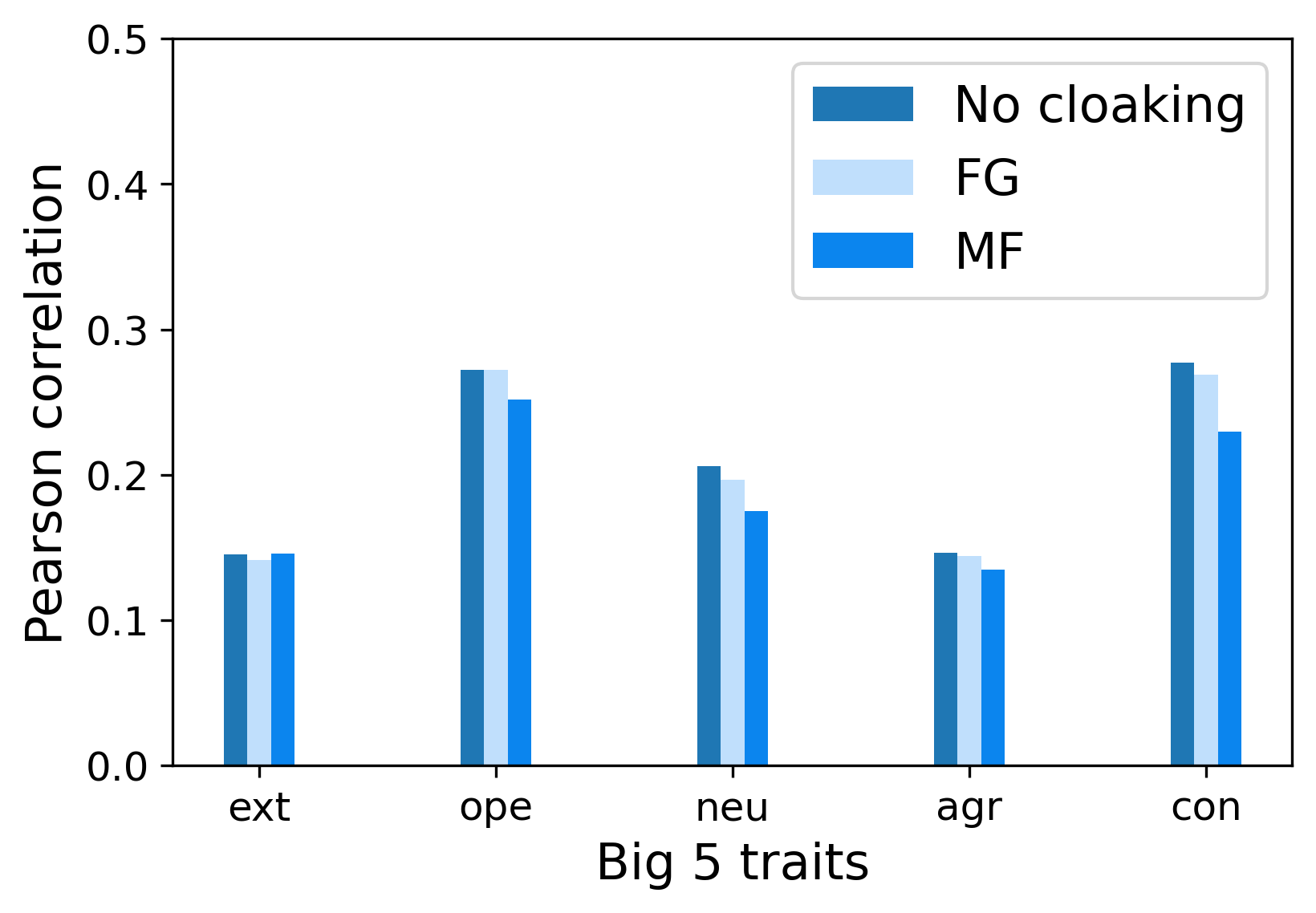}
         \caption{Democrat}
         \label{fig:y equals x}
     \end{subfigure}
     \hfill
     \begin{subfigure}[b]{0.47\textwidth}
         \centering
         \includegraphics[width=\textwidth]{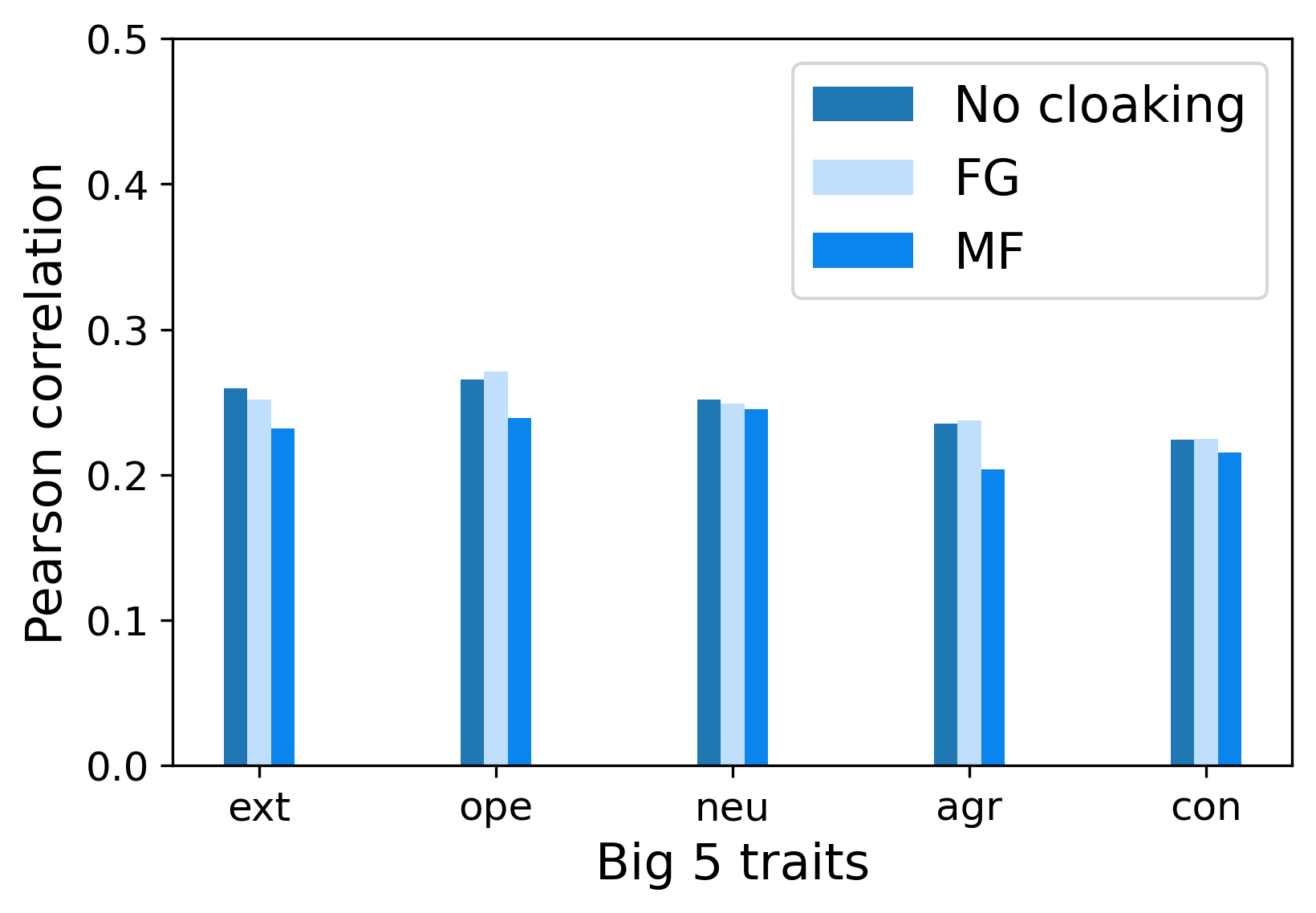}
         \caption{Republican}
         \label{fig:y equals x}
     \end{subfigure}
     \hfill
    \caption{Effect of cloaking on the predictive performance of the Big 5 traits.} \label{fig:exp2_results}
\end{figure}

We follow the set-up described in Section~\ref{subsec:set_up_exp2} to measure the impact of cloaking sensitive traits on the predictive performance of other prediction tasks (in this case the Big 5 traits).
Figure~\ref{fig:exp2_results} shows that the impact of cloaking metafeatures on the predictive performance of the Big 5 traits is larger on average than the impact of cloaking fine-grained features. For both strategies, the impact is largest when cloaking gender, followed by political orientation. 
The impact of both cloaking strategies on the predictive performance seems fairly \emph{small} in most cases, but we cannot truly judge the losses in value (corresponding to the small losses in predictive power) in a study such as this. 


\section{Discussion and Conclusion}

The digital traces we leave every day enable those who collect them to make intimate inferences about who we are. While such inferences might lead to desired personalization outcomes, they also pose a considerable threat to individuals' privacy and self-determination. In this paper, we examined the effectiveness and impact of two related privacy-enhancing cloaking strategies which conceal a portion of users' digital footprints to limit the ability of platforms to make predictions about underlying psychological traits. Although previous work has shown that such cloaking mechanisms can be effective in the short-term ~\citep{chen2017enhancing}, our findings suggest that their effectiveness rapidly declines over time. That is, as people continue to generate traces after the cloaking has been implemented, the system quickly relearns to draw those same inferences from the new data. We introduce a new cloaking strategy - one that is based on cloaking metafeatures rather than individual footprints  - and show how this strategy offers better longer-term privacy protection.

In addition, our findings also highlight potential trade-offs between privacy protection and personalized services. That is, while individuals might be interested in cloaking certain aspects of their identity (e.g., their sexual orientation), they might appreciate the benefits they receive from sharing other parts (e.g., their openness). We show that cloaking a particular trait likely has spillover effects on other traits that were not intentionally targeted. Although the trade-off between personalization and privacy is not a new idea, there are few (if any) empirical analyses of the actual trade-offs introduced by different privacy-enhancing techniques, including cloaking.\footnote{Prior work has shown a trade-off between privacy protection and advertising effectiveness ~\citep{goldfarb2011privacy}, but to our knowledge, not previously at the level of a specific prediction task.} 

The extent to which trading off personalization for enhanced privacy protection is desirable will depend on the specific context and preferences of the user. While some users might be willing to forsake targeted advertising for higher levels of privacy, others might favor convenience and service over the ability to conceal potentially unwanted aspects of their identity. The same is true for companies who might trade-off the ability to get highly granular consumer insights on all levels for a higher likelihood that consumers will stay on the platform and refrain from opting out of tracking and personalization altogether. We argue that different forms of cloaking can provide solutions that operate between the two extremes. On the one hand, they allow companies to keep collecting large amounts of data and monetize it within the boundaries set by users. On the other hand, users gain control over the level of personalization they feel comfortable with, while having the ability to inhibit unwanted inferences. 

\subsection{Practical Implications}

One of the main practical implications of this research is the extent to which cloaking can assist individuals in making smarter decisions when it comes to protecting their privacy. While data protection regulations such as the General Data Protection Regulation (GDPR) in Europe or the California Consumer Privacy Act (CCPA) have pushed for increased consumer control, there is a growing body of research suggesting that - without any support - individuals struggle to act as responsible stewards of their personal data. Research on the \emph{privacy paradox}, for example, reveals a stark discrepancy between a user's expressed concerns regarding online privacy and their actual behavior when sharing personal information. Despite expressing concerns about their privacy, individuals are often willing to share personal information online in exchange for personalized recommendations~\citep{barth2017privacy}. For instance, even though 93\% of USA citizens consider it important to maintain control over who can access their data, only a small fraction of people actually read the privacy policies of the services collecting their data~\citep{madden2015americans,solove2012introduction,matz2020privacy}.
One (obvious) reason why consumers do not succeed in achieving desired levels of privacy is their lack of knowledge about how their data is actually being collected and used~\citep{acquisti2020secrets}. Related is the \emph{acceptability gap}, which shows that users are more accepting of personalized services than of the collection of personal data required for these services~\citep{kozyreva2021public}. They overlook the relationship between them, and as a result, fail to engage in an adequate comparison of the value received from personalization to the value of keeping data private~\cite{kozyreva2021public}. As a consequence, most people tend to overvalue the short-term benefits of actions, such as using an app, over the long-term privacy risks, which are delayed and intangible~\citep{acquisti2004privacy}. 

Complicating matters further, research has suggested that people's apparent inaction regarding their privacy is also the result of them feeling that they have no control over the situation, and as a consequence simply give up (a phenomenon researchers have called \emph{digital resignation})~\citep{draper2019corporate,acquisti2020secrets}.  Finally, it may simply be that the perceived cost of protecting privacy is simply too high: either not using a service or possibly navigating an ultra-complicated web of documents and settings.  In all of these cases, providing transparency into how data is used and control over its use seems vital for consumer welfare and, in particular, for users to make informed privacy decisions~\citep{matz2020privacy}. Both privacy and transparency are essential prerequisites for establishing a trustworthy AI system~\citep{liu2022trustworthy}.

In this paper, we introduce a tool that could help guide individuals in making choices on their privacy settings online. Since the implications of sharing personal data are often difficult to anticipate, let alone trade-off for immediate convenience rewards, we need easy ways for people to move the dial between oversharing and undersharing. Cloaking offers such a lever and might encourage platforms to offer more mechanisms for users to control data-driven inferences and personalization (including targeted advertising) either through editing of the data items---the digital footprints---that are stored about them or through an explicit cloaking mechanism that hides footprints from the AI inference systems specifically.

Notably, our cloaking methodology depends on the cooperation of the platforms that collect this kind of data (like Facebook, Google, Spotify). While platforms might try to resist the introduction of technology that limits their ability to commercialize consumer insights, we argue that introducing certain levels of consumer control in the form of cloaking could eventually benefit them in the long-run. As stricter data protection regulations are introduced around the world - often empowering consumers to revoke access to their personal data - platforms might be forced to provide sufficient transparency and control in order to retain users and prevent them from option out of data collection entirely. Moreover, a gradual shift to higher levels of platform-driven user control might prevent legislators from introducing more paternalistic regulatory actions.

\section*{Acknowledgments}
This research was funded by Flemish Research Foundation (grant number 11N7723N). We would also like to thank Research Foundation-Flanders for their support for the research stay in New York (grant number V403523N).
Foster Provost thanks Ira Rennert and the Stern/Fubon Center for support.
This study was approved by the Institutional Review Board of the University of Antwerp (SHW2231, 7 February 2023).

\bibliographystyle{abbrvnat}  
\bibliography{references}  

\appendix


\newpage
\section{Results of other cloaking strategies} \label{subsec:results_other_strategies}
\subsection{Using domain-based metafeatures}
\begin{figure}[!htb]
    \centering
    \begin{subfigure}[b]{0.47\textwidth}
         \centering
         \includegraphics[width=\textwidth]{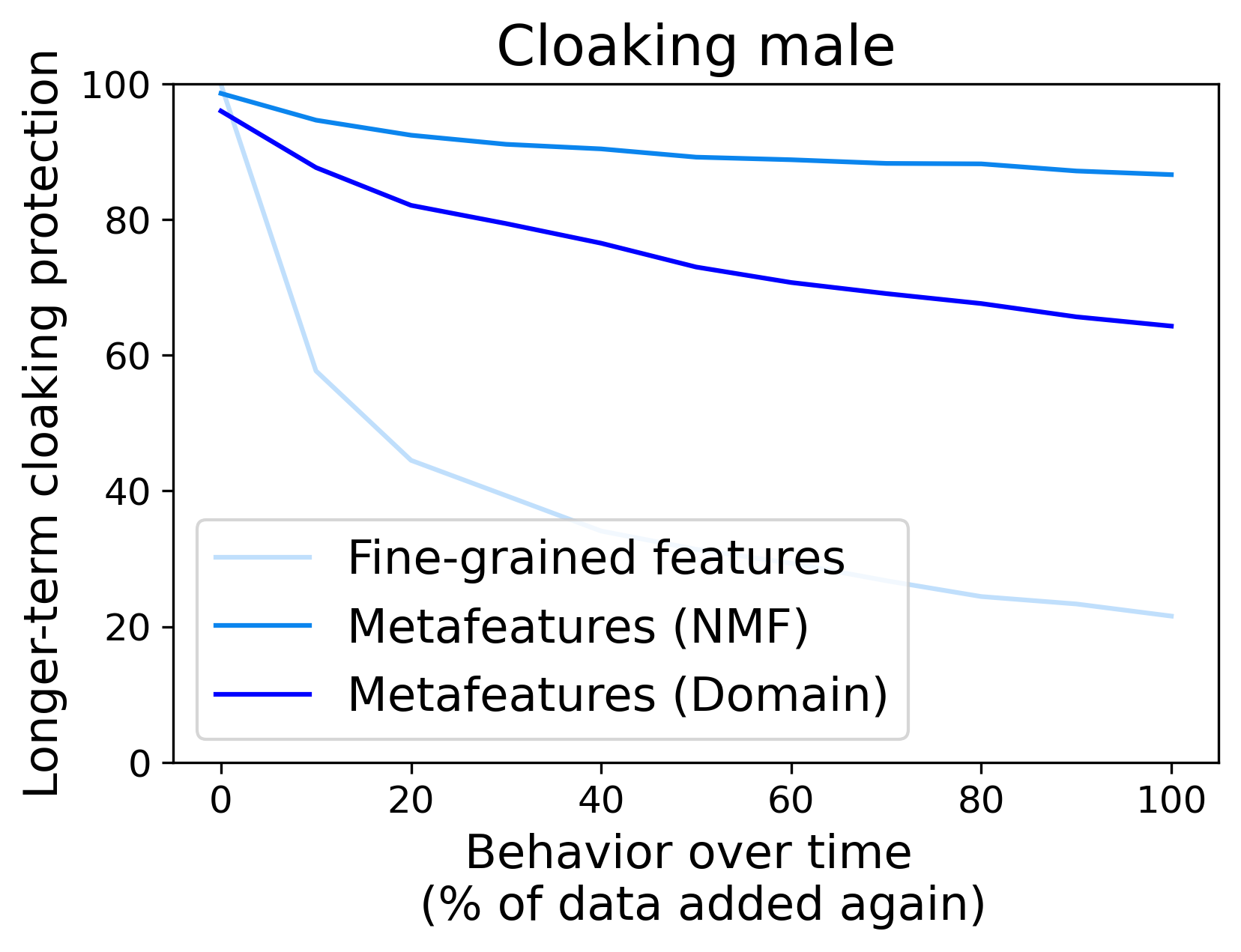}
     \end{subfigure}
     \hfill
     \begin{subfigure}[b]{0.47\textwidth}
         \centering
         \includegraphics[width=\textwidth]{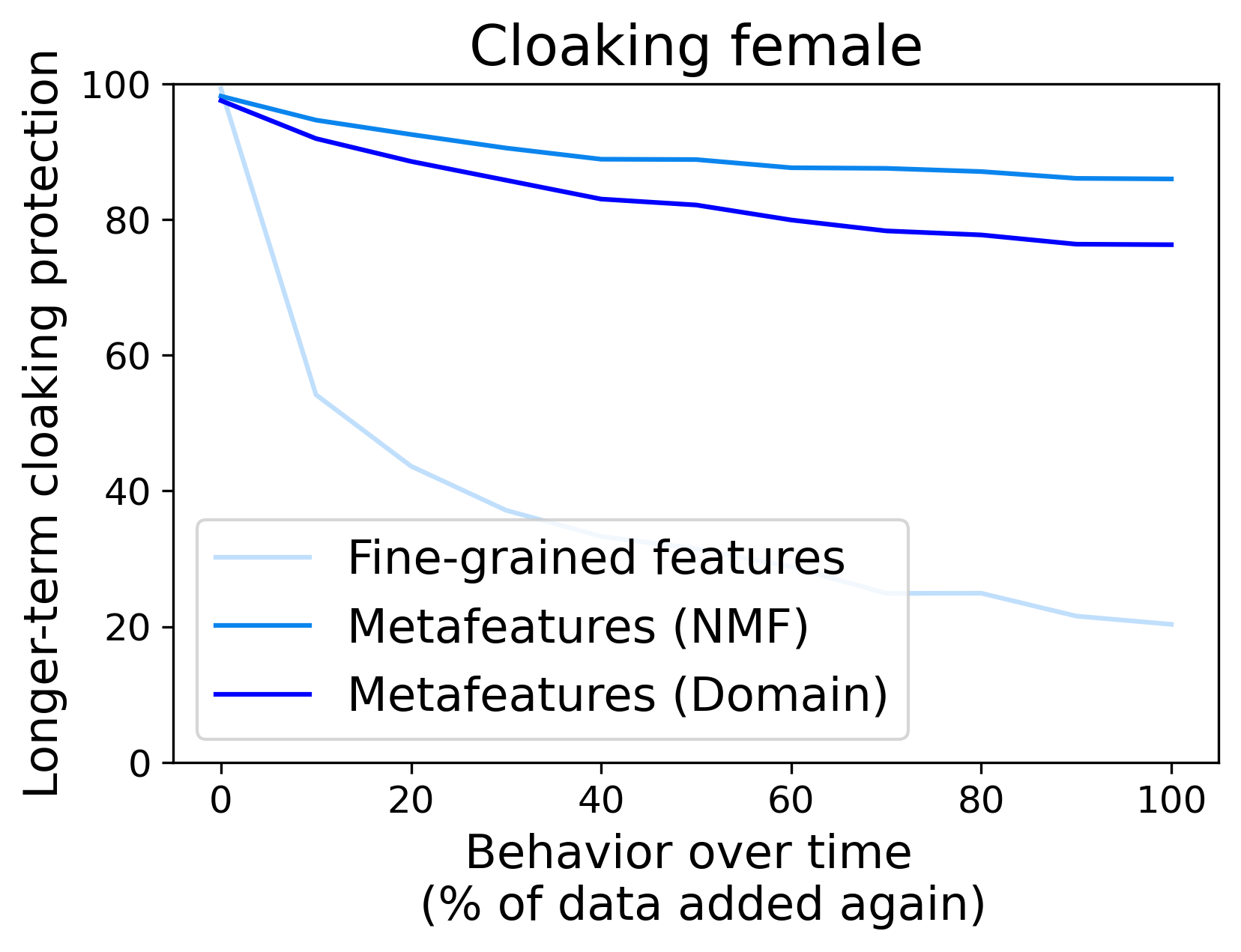}
     \end{subfigure}
     \hfill
     \begin{subfigure}[b]{0.47\textwidth}
         \centering
         \includegraphics[width=\textwidth]{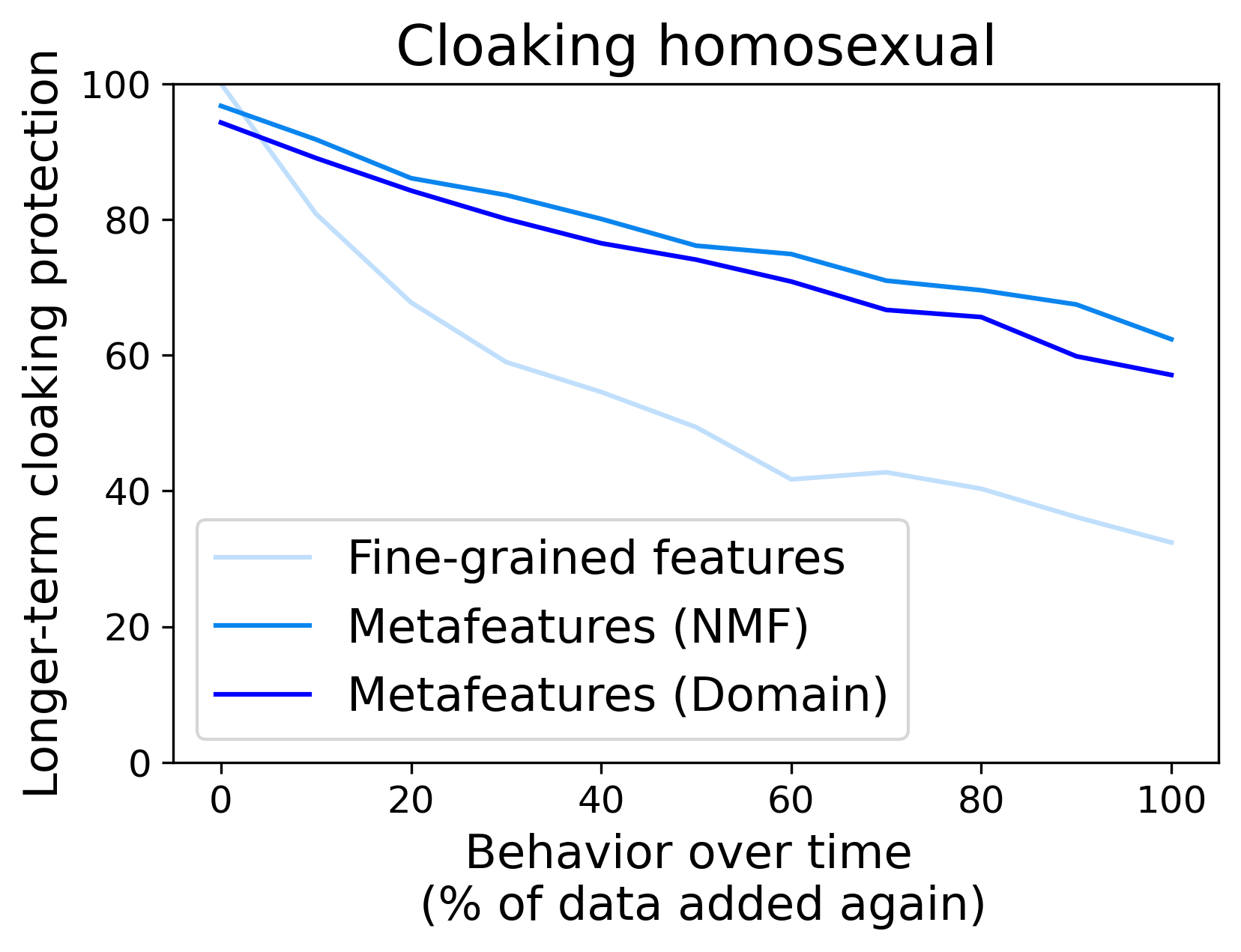}
     \end{subfigure}
     \hfill
     \begin{subfigure}[b]{0.47\textwidth}
         \centering
         \includegraphics[width=\textwidth]{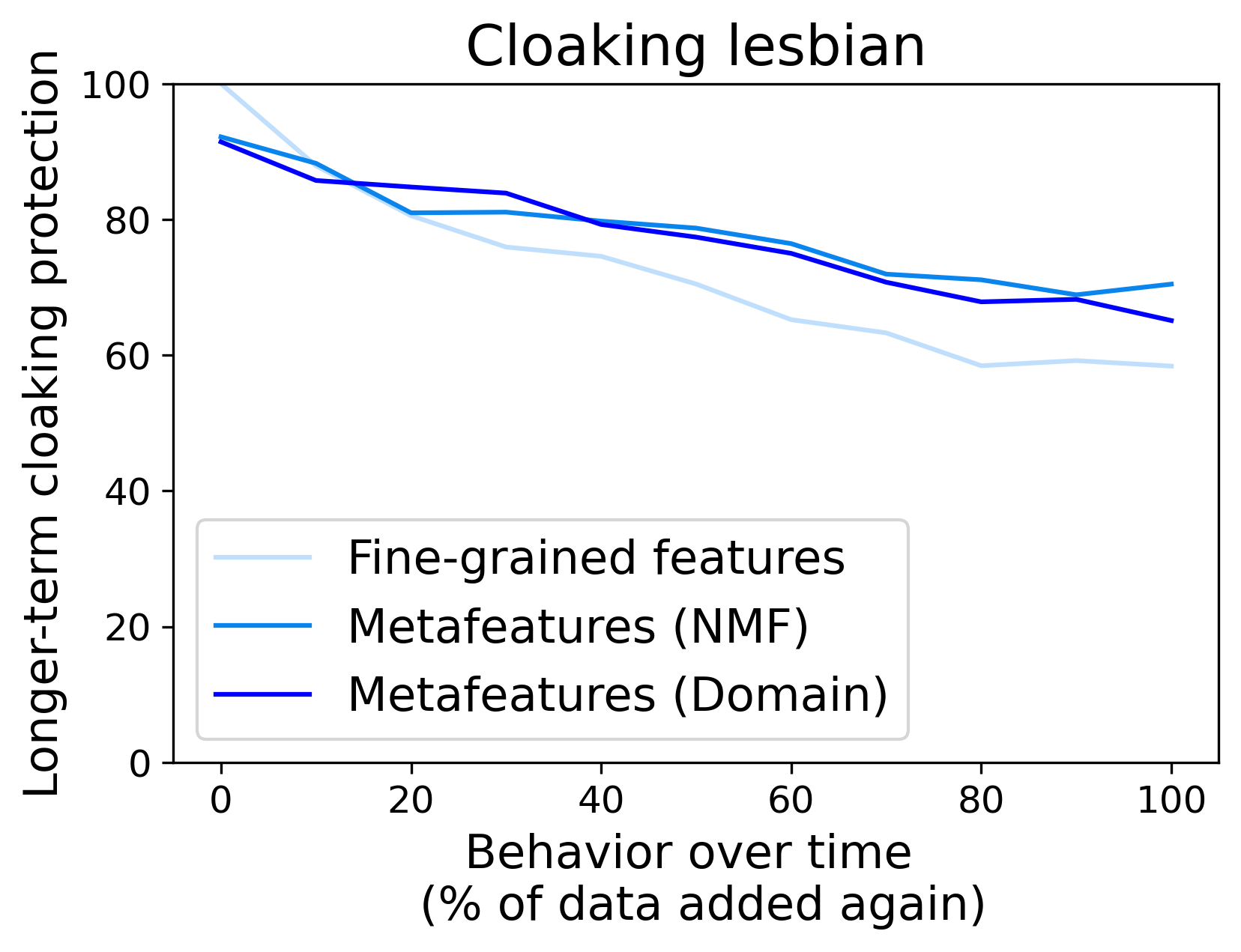}
     \end{subfigure}
     \hfill
     \begin{subfigure}[b]{0.47\textwidth}
         \centering
         \includegraphics[width=\textwidth]{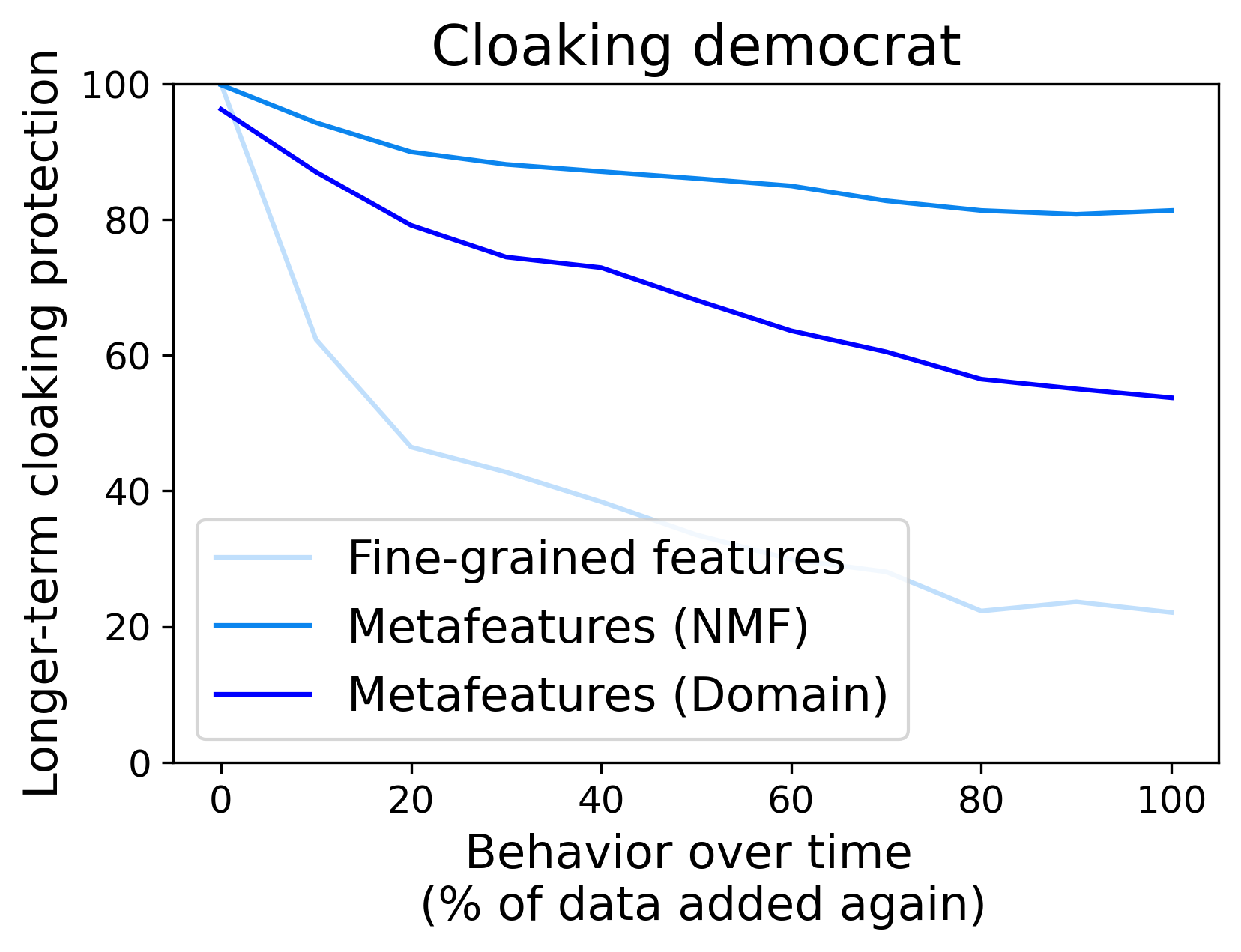}
     \end{subfigure}
     \hfill
     \begin{subfigure}[b]{0.47\textwidth}
         \centering
         \includegraphics[width=\textwidth]{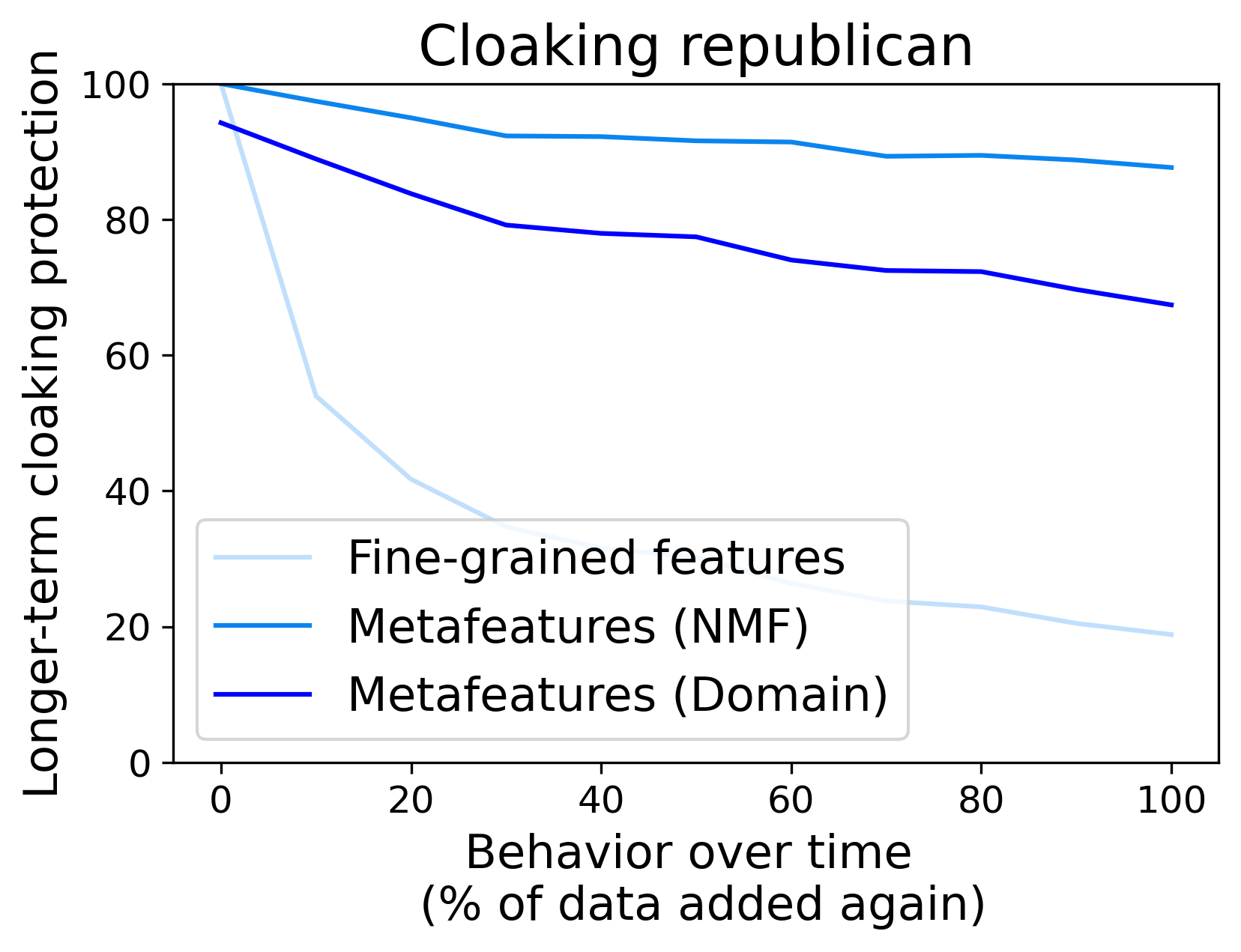}
     \end{subfigure}
     \hfill
    \caption{Longer-term cloaking protection over time when using domain-based metafeatures.} \label{fig:exp1_domain}
\end{figure}

\newpage
\subsection{Using explanations with a tolerance}
\begin{figure}[!htb]
    \centering
    \begin{subfigure}[b]{0.47\textwidth}
         \centering
         \includegraphics[width=\textwidth]{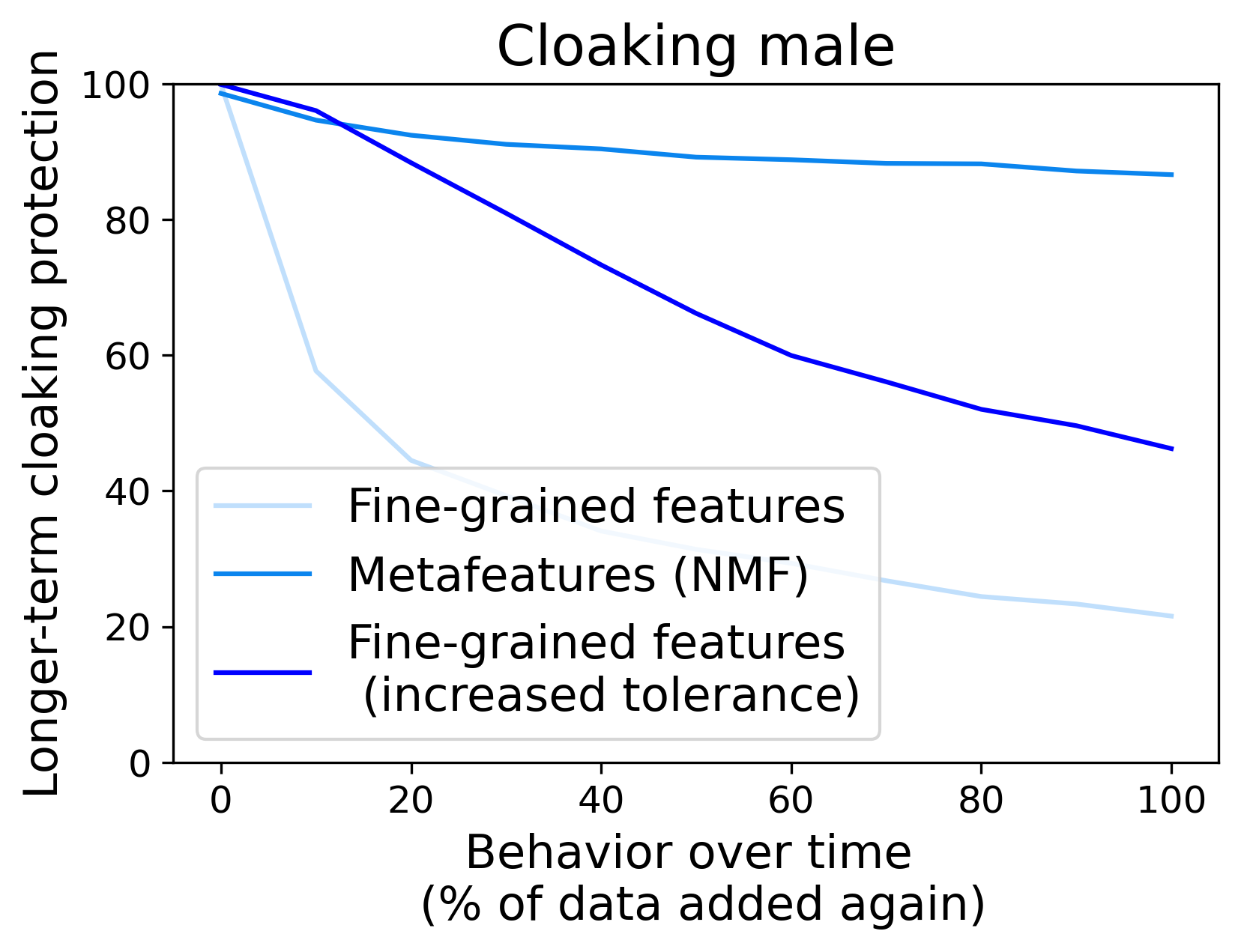}
         \label{fig:y equals x}
     \end{subfigure}
     \hfill
     \begin{subfigure}[b]{0.47\textwidth}
         \centering
         \includegraphics[width=\textwidth]{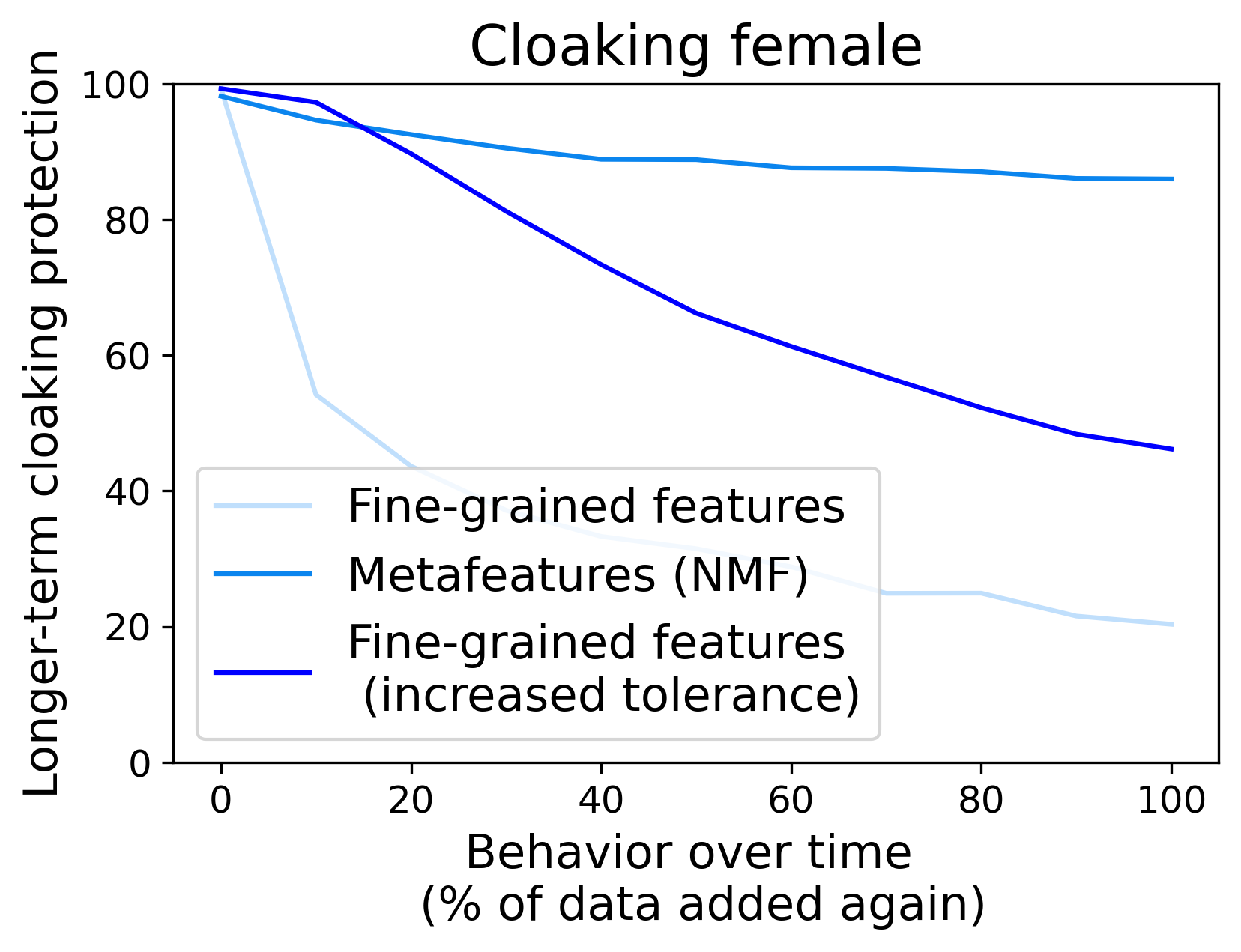}
         \label{fig:y equals x}
     \end{subfigure}
     \hfill
     \begin{subfigure}[b]{0.47\textwidth}
         \centering
         \includegraphics[width=\textwidth]{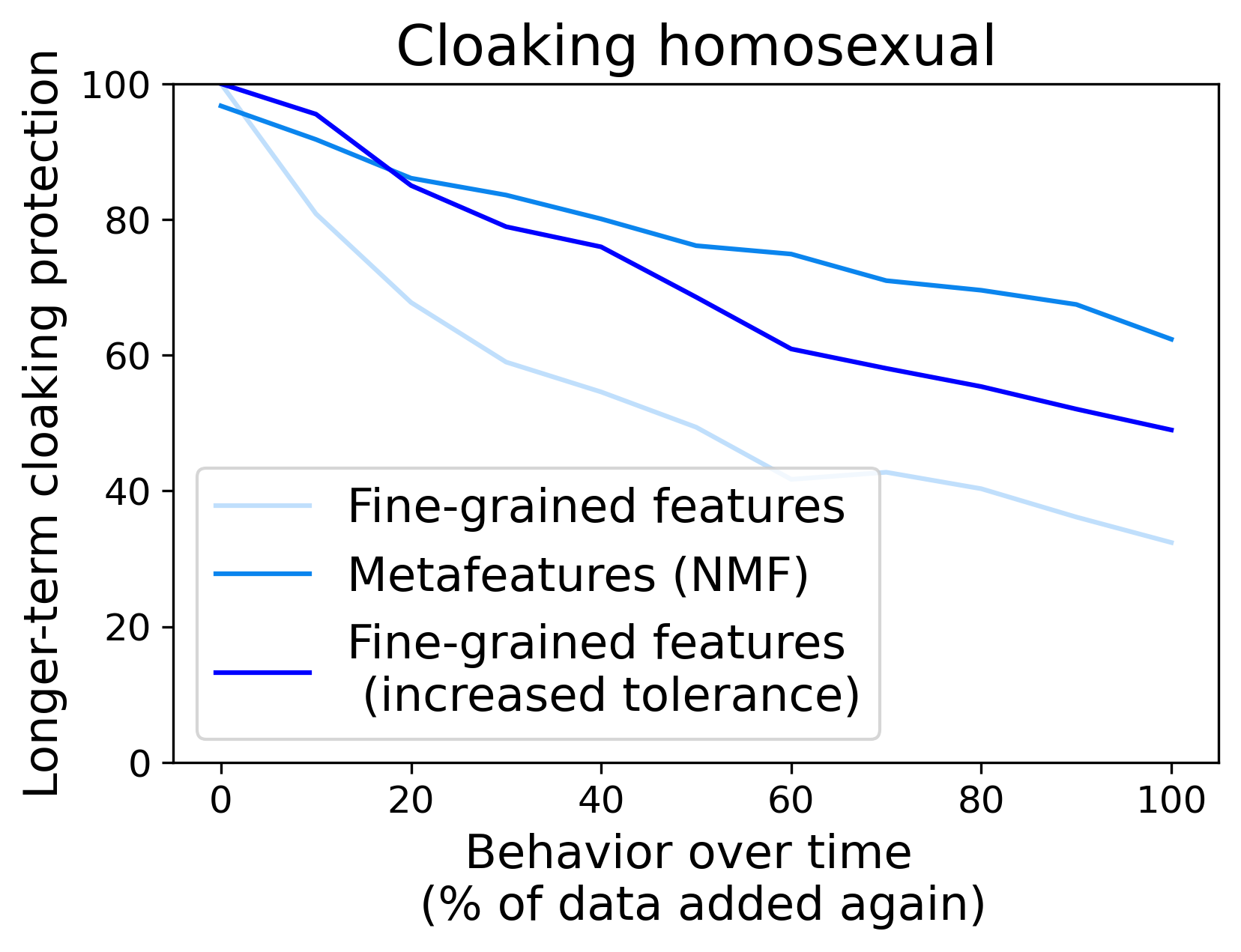}
         \label{fig:y equals x}
     \end{subfigure}
     \hfill
     \begin{subfigure}[b]{0.47\textwidth}
         \centering
         \includegraphics[width=\textwidth]{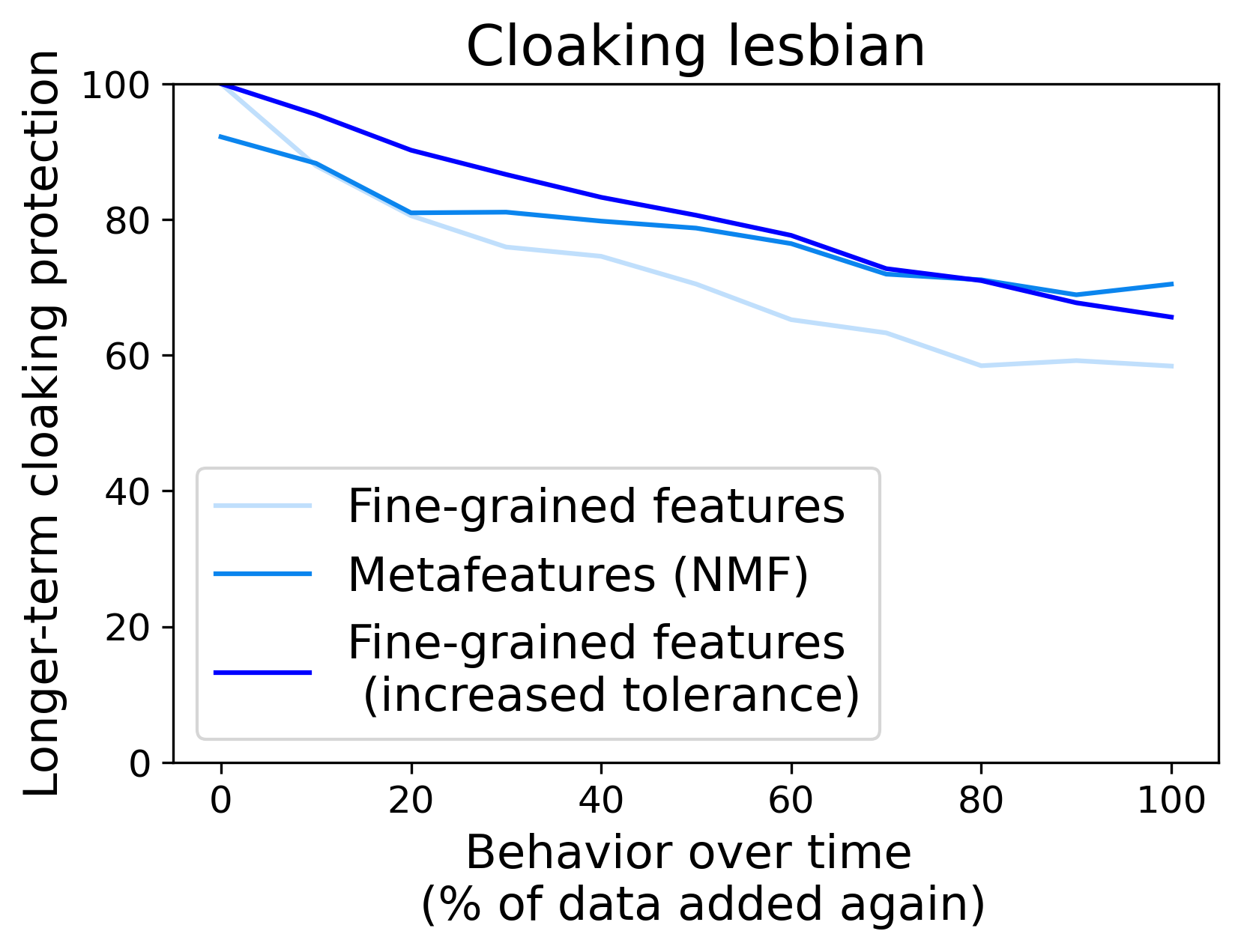}
         \label{fig:y equals x}
     \end{subfigure}
     \hfill
     \begin{subfigure}[b]{0.47\textwidth}
         \centering
         \includegraphics[width=\textwidth]{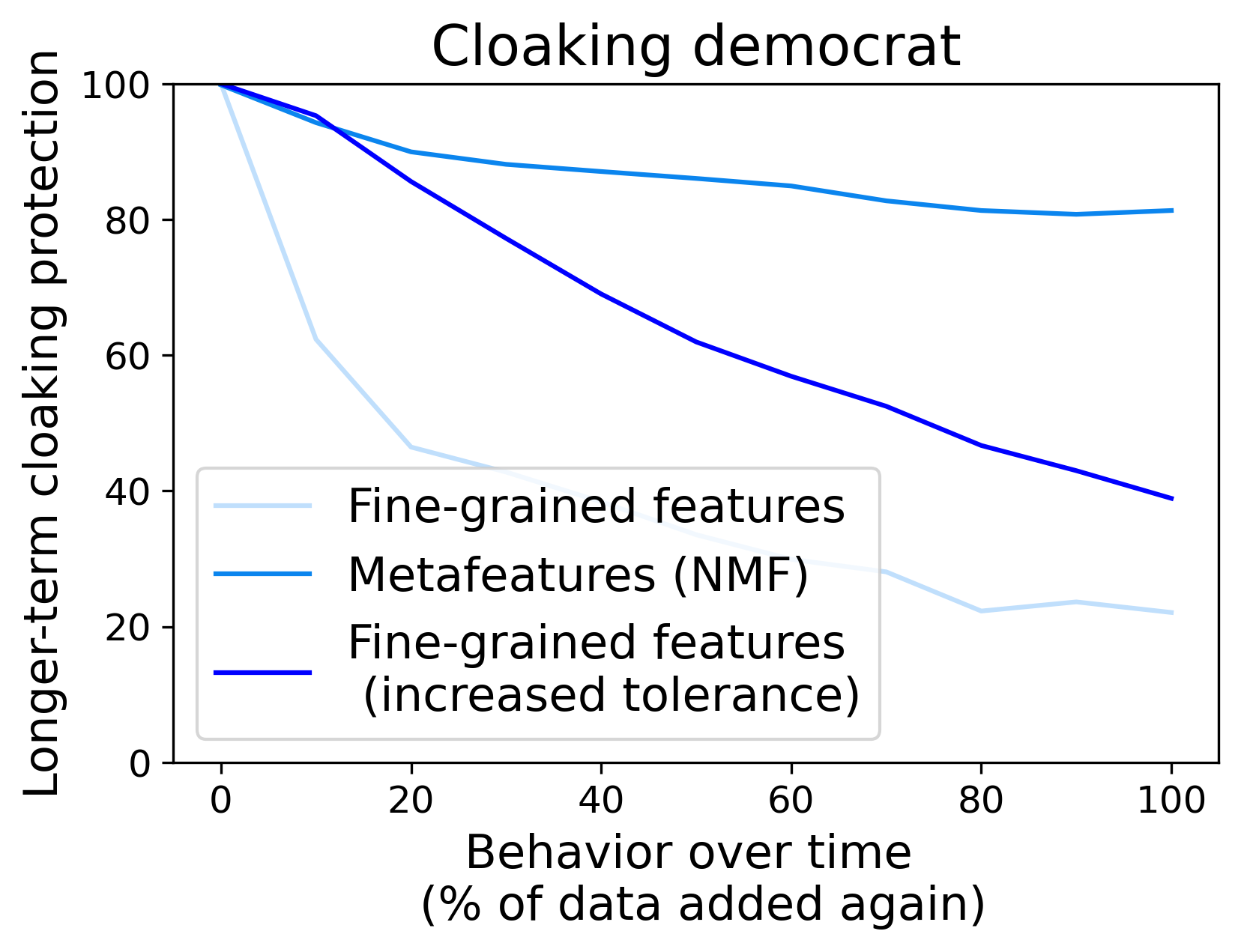}
         \label{fig:y equals x}
     \end{subfigure}
     \hfill
     \begin{subfigure}[b]{0.47\textwidth}
         \centering
         \includegraphics[width=\textwidth]{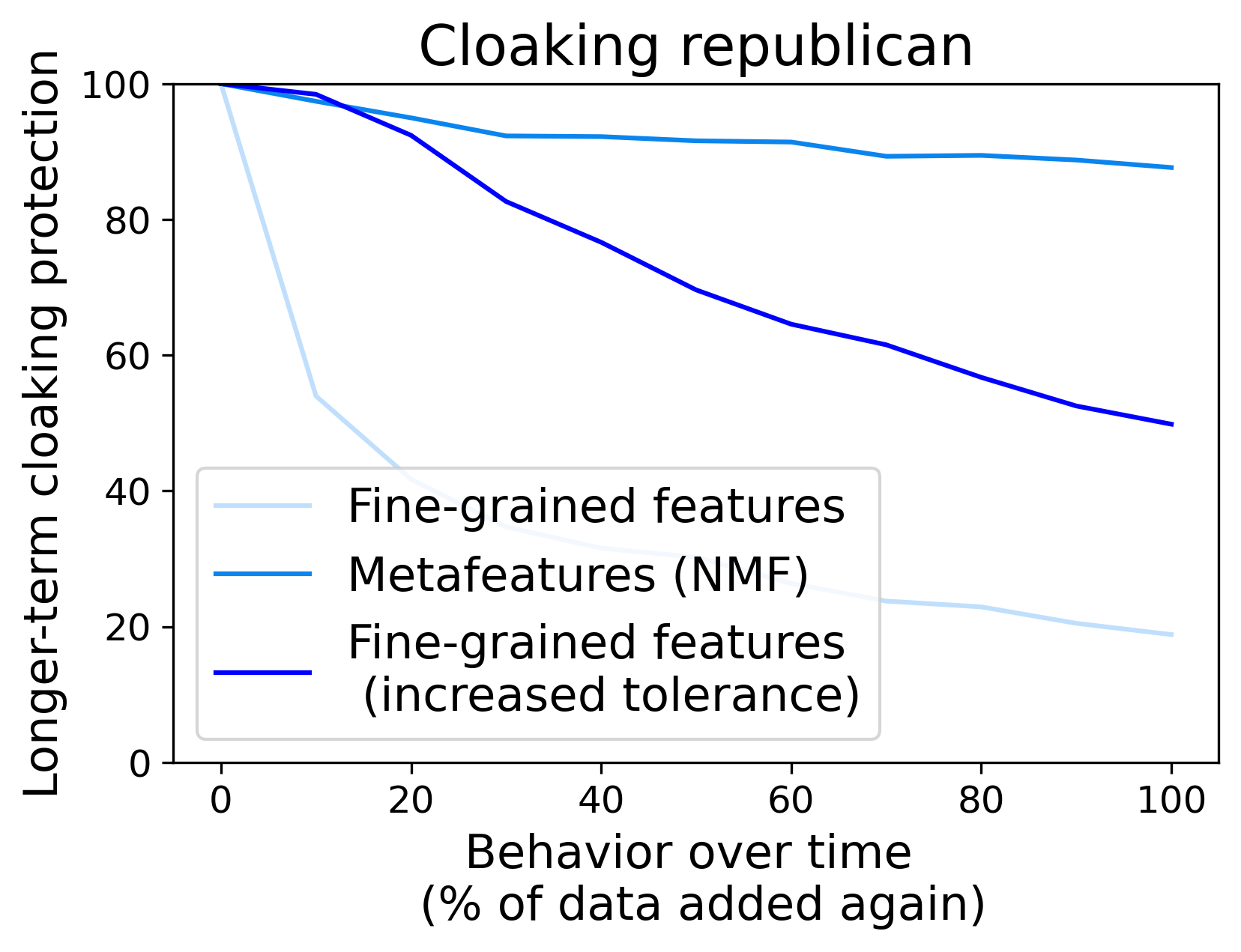}
         \label{fig:y equals x}
     \end{subfigure}
     \hfill
    \caption{Longer-term cloaking protection over time when using explanations with an additional tolerance level.} \label{fig:exp1_tolerance}
\end{figure}

\end{document}